\documentclass[
 reprint,
 superscriptaddress,
 amsmath,
 amssymb,
 aip,
 a4paper,
 floatfix,
 longbibliography,
 twocolumn
]{revtex4-2}

\usepackage{graphicx}
\usepackage{dcolumn}
\usepackage{bm}
\usepackage{mathrsfs}
\usepackage{physics}
\usepackage{xcolor}
\usepackage{microtype}
\usepackage[american]{babel}
\usepackage[utf8x]{inputenc}
\usepackage{dsfont}
\usepackage{siunitx} 
\usepackage{mathtools}
\usepackage{enumerate}

\newcommand{\ee}{\text{e}}
\newcommand{\ii}{\text{i}}

\newcommand\mustl{\stackrel{\mathclap{\normalfont\mbox{!}}}{\ll}}

\newcommand{\orcid}[1]{\href{https://orcid.org/#1}{\includegraphics[width=7pt]{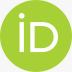}}}
\usepackage{soul}
\newcommand{\C}[1]{{\color{red}{#1}}}

\definecolor{cset-aps-blueberry}{RGB}{28,128,158}
\definecolor{cset-aps-blue}{RGB}{46,44,184}
\definecolor{cset-aps-turquoise}{RGB}{0,67,88}
\definecolor{cset-aps-limegreen}{RGB}{190,219,67}
\definecolor{cset-aps-green}{RGB}{31,138,112}
\definecolor{cset-aps-yellow}{RGB}{255,225,25}
\definecolor{cset-aps-orange}{RGB}{253,116,0}
\definecolor{cset-aps-red}{RGB}{219,0,43}
\definecolor{myred}{RGB}{255,0,20}
\usepackage{hyperref}
\hypersetup{%
    colorlinks=true,
    linkcolor={cset-aps-red},
    linkbordercolor={cset-aps-red},
    filecolor={cset-aps-orange},
    filebordercolor={cset-aps-orange},
    citecolor={cset-aps-blue},
    citebordercolor={cset-aps-blue},
    urlcolor={cset-aps-blueberry},
    urlbordercolor={cset-aps-blueberry},
    menucolor={cset-aps-limegreen},
    menubordercolor={cset-aps-limegreen},
    breaklinks=true,
    pdfborderstyle={/S/U/W 2},
    pdfpagemode=UseOutlines,
    pdfstartpage={1},
}
\usepackage{cleveref}

\AtBeginDocument{\RenewCommandCopy\qty\SI}

\begin{document}

\preprint{APS/123-QED}
\title{Time-dependent adiabatic elimination in matter-wave optics}
\collaboration{Published version available in  \href{https://doi.org/10.1116/5.0304521}{AVS Quantum Sci. \textbf{8}, 024403 (2026)};
licensed under a \href{http://creativecommons.org/licenses/by/4.0/}{Creative Commons Attribution [CC BY]} license.}
\author{Samuel Böhringer~\orcid{0009-0003-9835-4009}}%
\affiliation{Institut f{\"u}r Quantenphysik and Center for Integrated Quantum
    Science and Technology (IQST), Universit{\"a}t Ulm, Albert-Einstein-Allee 11, D-89081 Ulm, Germany}
\email{samuel.boehringer@uni-ulm.de}
\author{Alexander Bott~\orcid{0000-0002-7986-4834}}%
\affiliation{Institut f{\"u}r Quantenphysik and Center for Integrated Quantum
    Science and Technology (IQST), Universit{\"a}t Ulm, Albert-Einstein-Allee 11, D-89081 Ulm, Germany}
\email{alexander.bott@uni-ulm.de}
\author{Eric P. Glasbrenner~\orcid{0000-0002-0822-3888}}
\affiliation{Institut f{\"u}r Quantenphysik and Center for Integrated Quantum
    Science and Technology (IQST), Universit{\"a}t Ulm, Albert-Einstein-Allee 11, D-89081 Ulm, Germany}
\email{eric.glasbrenner@uni-ulm.de}
\date{\today}

\begin{abstract}
We show how the dynamics of a specific subset of states can be separated from the dynamic of the total quantum state via a time-dependent projector-based formalism of adiabatic elimination. 
Within our formalism, we assume an explicit time dependency in the coupling between both subsystems. 
Additionally, we do not assume that the elements of the Hamiltonian commute, as in matter-wave optics this is not given in general. Here, the center-of-mass degrees of freedom frequently need to be taken into account.
Assuming non-commutativity also enables the application of this formalism to matrices of arbitrary dimension and the treatment of the adiabatic elimination in high dimensional systems, where the elimination may be needed to reduce the overall complexity. In particular, we apply our formalism to typical mechanisms in matter-wave optics, Raman and Bragg diffraction.
\end{abstract}

\maketitle

\section{Introduction}
Quantum sensors that utilize the interaction of matter and light have emerged as powerful tools for precision measurements \cite{Schlippert2014, Freier2016Jun, Savoie2018Dec}.
Optical clocks have reached high precision in the measurement of time \cite{Ludlow2015Jun,Oelker2019Oct,McGrew2019Apr,Antoine2003Apr,Heavner2014} while atom interferometers are promising candidates for the search of dark matter and physics beyond the standard model \cite{Abend2024, Abdalla2025,Ufrecht2020Nov, DiPumpo2021nov, DiPumpo2022Apr, DiPumpo2023mar}.
In order to increase the precision of such experiments an in depth analysis of the center-of-mass (COM) motion is necessary.
In some applications the residual COM motion reduces the accuracy of measurement outcomes and an analysis of such effects can help to improve the trapping of atoms \cite{Dalibard1992Feb,Dalibard1985Nov,Dalibard1989Nov,Aspect1986Oct,Castin1991Apr,Bohringer2024}. 
\par
In reality atoms have multiple internal states and in many applications one is not interested in the evolution of the total quantum state but rather a specific subset that is utilized for measurements.
In such cases there are known methods to separate the evolution of the relevant subsystem from the irrelevant subsystem.
First of all there is the well known projector formalisms of Nakajima and Zwanzig that find application in open quantum systems \cite{Breuer2007,Nakajima1958Dec,Zwanzig1960Nov}. 
Closely connected to such a projector formalism is the adiabatic elimination that is frequently employed in quantum optics \cite{Scullyzubairy1997, schleich2011,Sanz2016,Bruhnke2024,Paulisch2014,Sanz2016,Semin2016,Glasbrenner2025,Bott2023}.
In particular, the work of Sanz et. al. in Ref.~\cite{Sanz2016} inspired us to generalize the projector approach of adiabatic elimination to explicit time dependency and to include the full COM motion. 
With our generalization we are able to derive effective Hamiltonians that contain explicit time dependencies in the couplings between the internal states as well as the individual COM motion of every state.
In particular, we derive the momentum dependency of the light shifts and the Rabi frequency together with its time dependency consistently without freezing any states in the irrelevant subsystem. 
\par
These effective Hamiltonians can be used to calculate the effective dynamics for the relevant subsystem only.
Furthermore, we explicitly show how our formalism reproduces well known results of Vitanov et. al. \cite{Vitanov2001} in the case where we do not consider the COM motion.
\par
The main goal of our work is to introduce a robust formalism that is applicable in matter-wave optics and allows to consistently consider COM motion of the states and also allows time-dependent couplings to the eliminated states.
\par
Our work is structured as follows: 
In section \ref{sec:projector:method}, we introduce the modified formalism of Ref.~\cite{Sanz2016} including time-dependent couplings of the states and arbitrary non-commuting operators. Subsequently, we show how this formalism connects to previous works shown in Refs.~\cite{Paulisch2014,Sanz2016,Torosov2009} in section~\ref{sec:connection}. Afterwards, we apply the formalism to a typical set up in matter-wave optics\cite{Giese2013,Giese2015,Hartmann2020,Li2024,Werner2024,Hartmann2020_2}, i.e. Raman diffraction, in section~\ref{sec:application}.
We explicitly simulate our resulting effective Hamiltonian for single Raman diffraction for the $\text{D}_2$ Line of ${}^{87}\text{Rb}$. 
We appended other examples of application in Appendices~\ref{appendix:pulse_shapes}~and~\ref{appendix:further_examples}.
Where we derive the effective Hamiltonian in closed form for typical pulse shapes e.g. box, sine squared and Blackman pulses.

\section{Time Dependent Projector Formalism}\label{sec:projector:method}

In this section, we present a generalization of the projector formalism for the adiabatic elimination of a irrelevant subsystem from a larger system. 
Our approach extends the method introduced by Sanz et al. in Ref.~\cite{Sanz2016} by allowing all terms in the Hamiltonian to be explicitly time-dependent. Similar to the method of Sanz et al.\cite{Sanz2016}, we do not assume that operators acting on different subsystems commute, enabling potential application to problems where COM degrees of freedom need to be taken into account, as is for example the case in the control of matter waves.
We therefore omit any special notation to indicate non-commutativity, as all operators are treated as non-commuting by default. 
The hat notation is reserved exclusively for the position and momentum operators, $\hat{x}$ and $\hat{p}$, respectively, or terms that are directly dependent on them and act on the COM degrees of freedom only.
\subsection{Time Evolution and Projector to Subspace}
Let us define the following unitary operator
\begin{align}
    U_{X}(t_0,t) = \mathcal{T}\exp(-\ii \int\limits_{t_0}^t\dd t^{\prime}X(t^{\prime})),
\end{align}
for a time-dependent hermitian operator $X\equiv X(t)$.
Here, $\mathcal{T}$ denotes the time-/path-ordering operator. 
By definition the unitary operator $U_{X}(t_0,t)$ satisfies the operator differential equations
\begin{align}
    \frac{\dd}{\dd t} U_{X}(t_0,t) &= -\ii X U_{X}(t_0,t)\\
    \frac{\dd}{\dd t} U_{X}^{\dagger}(t_0,t) &= \ii  U_{X}^{\dagger}(t_0,t)X.
\end{align}
Note, that the presence of the time-ordering operator $\mathcal{T}$ enforces a strict temporal ordering of $X(t^{\prime})$ under the integral.
As a result, when taking the time derivative of $U_{X}(t_0,t)$ and $U_{X}^{\dagger}(t_0,t)$, the operator $X(t)$ appears to the left or the right, respectively, in accordance with this ordering.
The notations and conventions in this section will be used consistently throughout the whole article.
With these definitions we are in the position to turn to our task at hand.
\par
As we want to separate the evolution of a relevant substate, that is a subdivision of the total quantum state, our first step is the division of the quantum state $\ket{\psi}$ and the Hamiltonian $H$ into two subspaces which we will call the relevant state space $\ket{\alpha}$ and irrelevant state space $\ket{\beta}$. 
Let us assume that we are only interested in the time evolution of the relevant state $\ket{\alpha}$ where also the population should be initialized.  
Therefore, we decompose the quantum state $\ket{\psi}$ into two subsystems by using the vector notation 
\begin{align}
\label{eq:def:state}
    \ket{\psi} = 
    \begin{pmatrix}
        \ket{\alpha}\\
        \ket{\beta}
    \end{pmatrix}.
\end{align}
In consequence, we express the Hamiltonian $H$ in that vector notation as \cite{Sanz2016}
\begin{align}
    \label{eq:def:hamiltonian}
    H =\hbar
    \begin{pmatrix}
        \Delta &\Omega^{\dagger}\\
        \Omega&\Xi
    \end{pmatrix},
\end{align}
where  $\Delta \equiv \Delta(t)$, $\Omega \equiv \Omega(t)$ and $\Xi \equiv \Xi(t)$ are time-dependent operators of dimensions matching the subsystems spanned by the states $\ket{\alpha}\bra{\alpha}$, $\ket{\beta}\bra{\alpha}$ and $\ket{\beta}\bra{\beta}$, respectively.

The time evolution of the total state $\ket{\psi} \equiv \ket{\psi(t)}$ is governed by the Schrödinger equation
\begin{align}
    \label{eq:schroedinger:equation}
    \ii \hbar \frac{\dd}{\dd t} \ket{\psi} = H \ket{\psi}.
\end{align}
When we insert the representation of the Hamiltonian Eq.~\eqref{eq:def:hamiltonian} and the quantum state in terms of the relevant and irrelevant substate, Eq.~\eqref{eq:def:state}, into the Schrödinger equation, we arrive at two coupled equations for each subspace \cite{Sanz2016}
\begin{subequations}
    \begin{align}
     \label{eq:schroedinger:equation:alpha}
        \ii\frac{\dd}{\dd t} \ket{\alpha} &= \Delta \ket{\alpha} + \Omega^{\dagger} \ket{\beta}\\
        \label{eq:schroedinger:equation:beta}
        \ii\frac{\dd}{\dd t} \ket{\beta} &= \Xi \ket{\beta} + \Omega \ket{\alpha}.   
    \end{align}
\end{subequations}
When we introduce a time-dependent projector \mbox{$P \equiv P(t)$} \cite{Sanz2016} with the property
\begin{align}
    P\ket{\alpha} &= \ket{\beta},
\end{align}
we are able to formally decouple Eqs.~(\ref{eq:schroedinger:equation:alpha})
and (\ref{eq:schroedinger:equation:beta}) leading to an equation for the relevant state $\ket{\alpha}$.

Thus, the Schrödinger equation, Eq.~\eqref{eq:schroedinger:equation}, can be written as \cite{Sanz2016}
\begin{subequations}
    \begin{align}
    \label{eq:schroedinger:equation:alpha_projected}
        \ii\frac{\dd}{\dd t} \ket{\alpha} &= ( \Delta  + \Omega^{\dagger} P)\ket{\alpha}\\
        \label{eq:schroedinger:equation:beta_projected}
        \ii\frac{\dd}{\dd t} (P\ket{\alpha}) &= \Xi P\ket{\alpha} + \Omega \ket{\alpha}  
    \end{align}
\end{subequations}

where we emphasize that the right-hand side of Eq.~\eqref{eq:schroedinger:equation:alpha_projected} can be understood as the effective Hamiltonian acting on the state $\ket{\alpha}$.
Therefore, we define the effective Hamiltonian \cite{Sanz2016}
\begin{align}
    H_{\alpha} = \Delta + \Omega^{\dagger}P
\end{align}
for the relevant state $\ket{\alpha}$.
This leaves us with the task to find $P$ in order to determine an explicit form for the effective Hamiltonian $H_{\alpha}$.

Combining Eq.~\eqref{eq:schroedinger:equation:alpha_projected} and \eqref{eq:schroedinger:equation:beta_projected} yields an non-linear operator differential equation of Riccati type
\begin{align}
    \ii\frac{\dd}{\dd t}P = \Xi P - P \Delta + \Omega -P\Omega^{\dagger} P.\label{eq:projector_differentialequation}
\end{align}
for the projector $P$ \cite{Bruhnke2024}.
When we assume that the total population is initially in the relevant state $\ket{\alpha}$ at the initial time $t_0$, then the initial state reads  
\begin{align}
    \ket{\psi(t_0)} =
\begin{pmatrix}
        \ket{\alpha(t_0)}\\
        0
    \end{pmatrix}
\end{align}
where the initial population in the irrelevant subsystem is $\ket{\beta(t_0)} = 0$.
Hence, the initial condition for the projector is given by 
\begin{align}
    \label{eq:projector_initial}
    P(t_0) = 0
\end{align}
where we note that the zero on the right-hand side should be understood as an operator with matching dimensions that yields zero if applied to $\ket{\beta}$.

Using the initial condition, Eq.\eqref{eq:projector_initial}, together with the operator differential equation, Eq.~\eqref{eq:projector_differentialequation}, we are now in a position to derive an approximate solution for the projector $P$. 

Since we eliminate the irrelevant state $\ket{\beta}$ from our description, this procedure is justified only when the coupling between the relevant and irrelevant sectors is small compared to the minimum spectral gap separating them \cite{Sanz2016}. 
Consequently, we define the minimal gap in the spectra of $\Xi$, $\sigma(\Xi)$, and $\Delta$ with its spectrum $\sigma(\Delta)$ via
\begin{align}
\gamma_{\star} = \min\limits_{\delta\in\sigma(\Delta),\xi \in\sigma(\Xi)} \abs{\xi - \delta}.
\end{align}

Now we consider the evolution of the projector on the timescale
\begin{equation}
	\tau = \gamma_{\star}t.
\end{equation}
Then Eq.~\eqref{eq:projector_differentialequation} takes the form
\begin{align}
    \ii\frac{\dd}{\dd \tau}P = \Xi_{\star} P -  P \Delta_{\star} + \frac{1}{\gamma_{\star}}\Omega -\frac{1}{\gamma_{\star}}P\Omega^{\dagger} P
\end{align}
where the operators $\Xi_{\star}$ and $\Delta_{\star}$ are rescaled with respect to the energy gap $\gamma_{\star}$.
We assume that the minimal energy gap $\gamma_{\star}$ is larger than the coupling $\Omega$ but in the same order of magnitude as the first term, such that all terms containing the coupling are suppressed with $1/\gamma_{\star}$.
With these properties in mind we assume the series expansion 
\begin{align}
    P = \sum\limits_{\ell= 0}^{\infty}\frac{1}{\gamma_{\star}^{\ell}}P_\ell.
\end{align}
for the projector $P$ \cite{Sanz2016}.

Inserting the expansion, matching orders in $1/\gamma_{\star}$ and reverting the time transformation, we find a recursive operator differential equation that is of the form
\begin{align}
    \ii\frac{\dd}{\dd t}P_{\ell} =  \Xi P_{\ell} - P_{\ell} \Delta + \mathcal{F}_{\ell}[P_{<\ell}]\label{eq:operator_equation}
\end{align}
where $\mathcal{F}[P_{<\ell}]$ may depend on all projectors of orders lower than $\ell$.
For the inhomogeneity $\mathcal{F}[P_{<\ell}]$, we find
\begin{align}
    \mathcal{F}_{0}[P_{<0}] &= 0\\
    \mathcal{F}_{1}[P_{<1}] &= \Omega - P_0\Omega^{\dagger}P_0\\
    \mathcal{F}_{\ell}[P_{<\ell}] &= - \sum\limits_{k=0}^{\ell-1} P_{\ell-1-k}\Omega^{\dagger}P_k \text{ for }\ell>1.
\end{align}
Eq.~\eqref{eq:operator_equation} is an inhomogeneous Sylvester Equation for which the solution is given by \cite{Behr2019Dec,fausett2009sylvester,Bruhnke2024}
\begin{align}
    \begin{split}
        P_{\ell} =& U_{\Xi}(t_0,t)P_{\ell}(t_0)U^{\dagger}_{\Delta}(t_0,t) \\&- \ii \int\limits_{t_0}^t \dd s \,U_{\Xi}(s,t)\mathcal{F}_{\ell}[P_{<\ell}]  U^{\dagger}_{\Delta}(s,t).
    \end{split}
\end{align}
With the initial condition $P_{\ell}(t_0) = 0$, we find 
\begin{align}
    P_{\ell} = - \ii \int\limits_{t_0}^t \dd s \,U_{\Xi}(s,t)\mathcal{F}_{\ell}[P_{<\ell}]  U^{\dagger}_{\Delta}(s,t)\label{eq:projector_solution}
\end{align}
and consequently 
\begin{align}
    \label{eq:P:0}
    P_0(t) =& 0\\
    P_1(t)=& - \ii \int\limits_{t_0}^t \dd s \,U_{\Xi}(s,t)\Omega(s) U^{\dagger}_{\Delta}(s,t)\label{eq:P:1}
\end{align}
for the first two orders. 

Note that Eq.~\eqref{eq:projector_solution} can be interpreted as contour ordered propagation of $\mathcal{F}[P_{<\ell}]$ via the Hamiltonians $\Xi$ and $\Delta$. 
There are also other methods available for the perturbative analysis of such expressions \cite{Ufrecht2020,Keldysh,Stefanucci2013}.
This projector is very closely connected to the non-operator case presented in Ref.~\cite{Paulisch2014} as can be seen from the structure of the integral in Eq.~\eqref{eq:projector_solution}.

\subsection{Conditions for validity of the the Adiabatic Elimination}
We need the following properties to apply the adiabatic approximation:
\begin{enumerate}[i)]
    \item The minimal energy gap $\gamma_{\star}$ between both manifolds needs to be larger than the coupling $\Omega$.
    \item The minimal energy gap $\gamma_{\star}$ between both manifolds needs to be larger than the level splitting within the respective manifolds.
    \item The time dependency of the coupling $\Omega$ needs to be smooth enough such that the coupling does not change considerable on the time scale $\pi/\gamma_{\star}$. This implies an upper time limit ${\pi \norm{\dd \Omega(s)/\dd s}(t-t_0)/\gamma_{\star}\ll 1}$.
\end{enumerate}
Only if these conditions are met by the system and the initial states, our formalism remains valid and is a good approximation.
A brief motivation for those conditions can be found in Appendix~\ref{sec:proof}.
In practice the minimal energy gap is often the frequency difference between the relevant lower energy states to the ancilla manifold. This gap is usually in the optical range and therefore at the order of hundreds of $\text{THz}$, while the energy gap within the manifolds is usually at the order of several $\text{GHz}$ \cite{steck2000sodium}, with single photon Rabi frequencies in the range of several $\text{MHz}$ \cite{Gentile1989Nov} up to the $\text{GHz}$ range \cite{Beterov_2024}  but typical pulse times are in the order of several \textmu$\text{s}$ \cite{HU2023170637,Hartmann2020_2}. This indicates that in most cases the single photon Rabi frequency is still magnitudes lower than the energy gap between the two manifolds. Hence, most experiments that use typical optical transitions in atoms fulfill those conditions.

\subsection{Effective Hamiltonian}
The effective Hamiltonian up to order $N$ in our expansion with the projectors is then given by 
\begin{align}
    \label{eq:effecfive_hamiltonian}
    H_{\alpha,N} = \hbar\Delta + \hbar\Omega^{\dagger}\sum\limits_{\ell=0}^{N}P_{\ell}.
\end{align}
Eq.~\eqref{eq:effecfive_hamiltonian} together with Eq.~\eqref{eq:projector_solution} are the main results of this section and allow us to calculate an effective Hamiltonian for the relevant state $\ket{\alpha}$ up to a given order $N$ for a time-dependent Hamiltonian as long as the coupling between the relevant and irrelevant state $\Omega$ is small.
The Hamiltonian in this framework is not necessarily hermitian.
However, there are methods to address this issue like presented in Ref.~\cite{Sanz2016}.
We will refrain from going into detail here, as we are interested only in the leading order term.

\section{Connection to other methods and Markovian approximation}\label{sec:connection}
In this section we show how our formalism connects to the results of Ref.~\cite{Paulisch2014} and Ref.~\cite{Sanz2016} in the limit of no explicit time dependency. We demonstrate that our approach is in fact a non-Markovian approximation where we assume that the state has memory described by the homogeneous solution of the Schrödinger equation of the relevant states.

Afterwards we show that our formalism applied in a suitable interaction picture leads to the same integral formula as derived in Ref.~\cite{Torosov2009}. The difference here is that our approach is not limited to commutativity or specific system sizes.

\subsection{The limit of no time dependency}
In the limit where none of the operators $\Delta,\Xi$ or $\Omega$ are time dependent, the formal integral solution for the state $\ket{\beta(t)}$ as determined by the Schrödinger equation, Eq.~\eqref{eq:schroedinger:equation:beta}, is given by \cite{Paulisch2014,Torosov2009}
\begin{align}
 \ket{\beta(t)} = - \ii \int\limits_{t_0}^t \dd s \,U_{\Xi}(s,t)\Omega \ket{\alpha(s)}.\label{eq:markov_integral}
\end{align}
The typical Markov approximation corresponds to eliminating the memory of the state $\ket{\alpha(s)}$ by the replacement $\ket{\alpha(s)}\rightarrow \ket{\alpha(t)}$ \cite{Paulisch2014}.
After applying this elimination the effective Hamiltonian for $\ket{\alpha}$ can be determined by performing the integral in Eq.~\eqref{eq:markov_integral} and dropping terms that oscillate with $\Xi$ \cite{Paulisch2014}. We find the lowest order adiabatic Hamiltonian
\begin{align}
    H_{\text{Markov}} = \Delta - \Omega^{\dagger} \Xi^{-1}\Omega
\end{align}

with this procedure.
Ref.~\cite{Paulisch2014} expands upon this by inserting a memory via ${\ket{\alpha(s)}\approx \ket{\alpha(t)}-(t-s)\frac{\dd}{\dd t}\ket{\alpha(t)}}$ \cite{Paulisch2014}.

In consequence the resulting Hamiltonian takes the form \cite{Paulisch2014}
\begin{align}
    H_{\text{Paulisch}} = \left(1+\Omega^{\dagger}\Xi^{-2}\Omega\right)^{-1}\left( \Delta - \Omega^{\dagger} \Xi^{-1}\Omega \right)\label{eq:Hamiltonian_Paulisch},
\end{align}

modifying the Markovian Hamiltonian $H_{\text{Markov}}$ by the factor $ \left(1+\Omega^{\dagger}\Xi^{-2}\Omega\right)^{-1}$.

Ref.~\cite{Sanz2016} instead proposes to analytically solve the projector equation Eq.~\eqref{eq:projector_differentialequation} for the limit of a time independent projector. This leads to the purely algebraic equation 
\begin{align}
    0 = \Xi P - P \Delta + \Omega -P\Omega^{\dagger} P.\label{eq:projector_equation_Sanz}
\end{align}
After a series expansion and by forcing a unitary time evolution the effective Hamiltonian takes the form \cite{Sanz2016}
\begin{align}
    H_{\text{Sanz}} = \Delta - \Omega^{\dagger}\Xi^{-1}\Omega -\frac{1}{2}\left(\Omega^{\dagger}\Xi^{-2}\Omega\Delta+\Delta\Omega^{\dagger}\Xi^{-2}\Omega\right) \label{eq:Hamiltonain_Sanz},
\end{align}

adding the term $-\frac{1}{2}\left(\Omega^{\dagger}\Xi^{-2}\Omega\Delta+\Delta\Omega^{\dagger}\Xi^{-2}\Omega\right)$ to the Markovian Hamiltonian $H_{\text{Markov}}$ instead of an overall factor. In the case of commuting operators $\Omega$, $\Xi$ and $\Delta$, this Hamiltonian results from an expansion of the perturbative factor in $H_{\text{Paulisch}}$.

Examining our result for the projector in first order, Eq.~\eqref{eq:P:1}, it becomes clear that our method predicts the evolution to follow the homogeneous part of Eq.~\eqref{eq:schroedinger:equation:alpha} via ${\ket{\alpha(s)}\approx U^{\dagger}_{\Delta}(s,t)\ket{\alpha(t)}}$ to lowest order. A physical argument for this is that we assume that the coupling $\Omega$ is small and will only slightly modify the homogeneous solution. The latter therefore describes the lowest-order behavior of the system in our approach, which as a result is set apart from the work of Ref.~\cite{Paulisch2014}.

In comparison to Ref.~\cite{Sanz2016}, we include a possible time-dependence of the Hamiltonian, which becomes evident from the comparison of Eq.~\eqref{eq:projector_differentialequation} with Eq.~\eqref{eq:projector_equation_Sanz}. The operator differential equation \eqref{eq:projector_differentialequation} from our approach includes the time-dependency of the Hamiltonian via the time-derivative of the projector $P$. Since the Hamiltonian in the formalism from Ref.~\cite{Sanz2016} is assumed to be time-independent, the derivative of $P$ is absent from Eq.~\eqref{eq:projector_equation_Sanz}.

The structure of the Hamiltonian resulting from our approach is also slightly different from Ref.~\cite{Sanz2016} as we find
\begin{align}
    H_{\alpha,1} = \Delta - \ii \Omega^{\dagger}\int\limits_{t_0}^t \dd s \,U_{\Xi}(s,t)\Omega U^{\dagger}_{\Delta}(s,t)\label{eq:Hamiltonian_Sam}
\end{align}
by inserting Eq.~\eqref{eq:P:1} into Eq.~\eqref{eq:effecfive_hamiltonian}.

The integration is not trivial due to multiple multiplied matrices, making a direct comparison with the other Hamiltonians complicated.
When we limit ourselves to the case of diagonal and commuting $\Xi$ and $\Delta$, we do however find the effective Hamiltonian
\begin{align}
    H_{\alpha,1} = \Delta - \sum\limits_{\ell,k,m}\frac{\Omega^{\dagger}_{\ell k}\Omega_{k m}}{\Xi_{kk}-\Delta_{mm}}\ketbra{\ell}{m}
\end{align}
after dropping the quickly oscillating terms. This Hamiltonian is similar to the Markovian Hamiltonian $H_{\text{Markov}}$ apart from the inclusion of $\Delta_{mm}$ in the denominator of the second term. Factoring out $\Xi_{kk}$ in the denominator and expanding the second term for small $\Delta_{mm}/\Xi_{kk}$ yields the Hamiltonian $H_{\text{Sanz}}$, which itself is an expansion of $H_{\text{Paulisch}}$ for this limiting case. Therefore the results for all three approaches merge into the same Hamiltonian for diagonal and commuting elements and in the limit of a large detuning $\Xi$ of the Hamiltonian.

\begin{figure}
    \centering
    \includegraphics[width=
    \columnwidth]{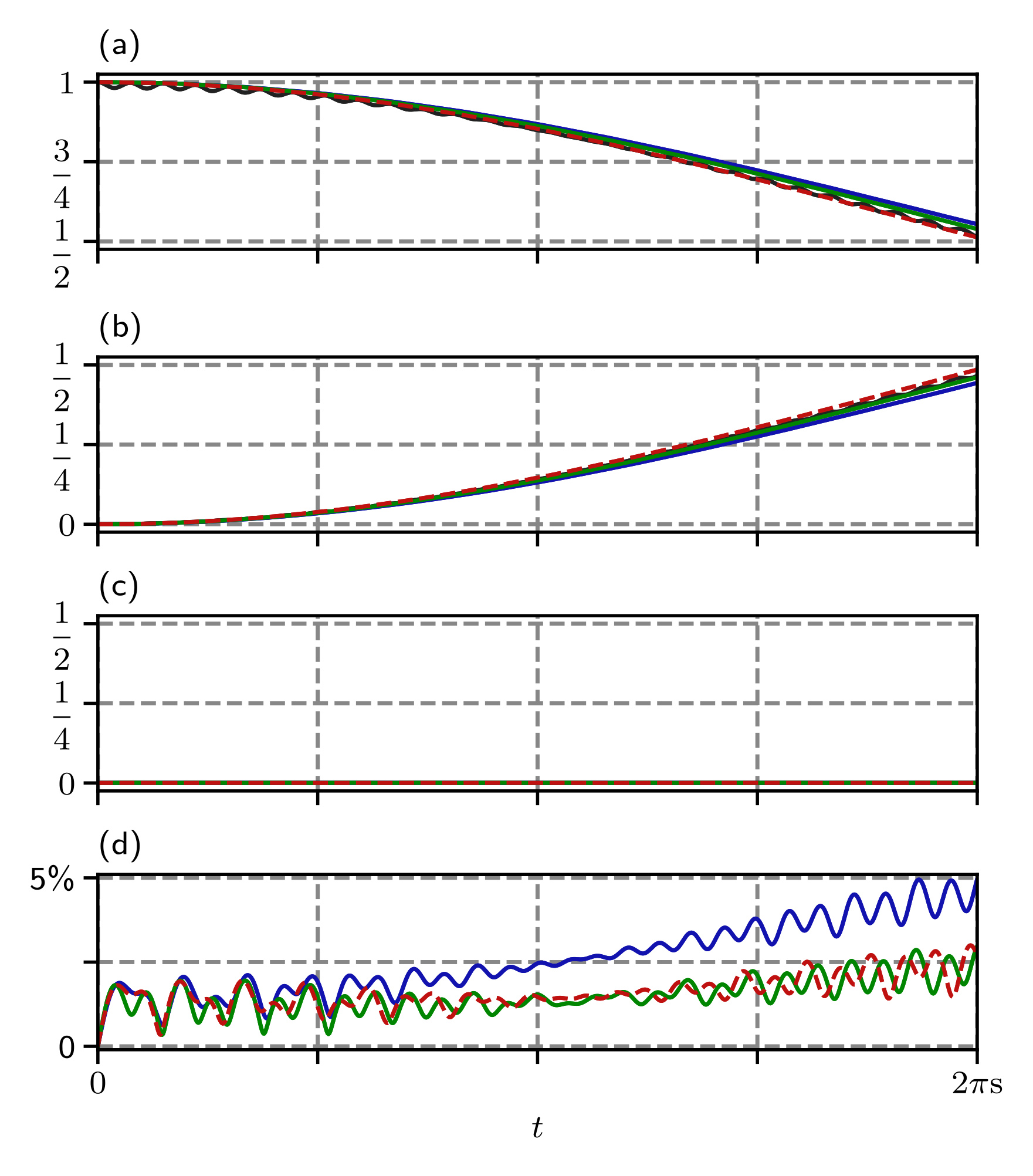}
    \caption{Comparison of methods for the adiabatic elimination of two ancilla states in a five level system. (a), (b) and (c) shows the population of the three states in the relevant states $\ket{\text{g}}$, $\ket{\text{m}}$ and $\ket{\text{e}}$ respectively. In black the numerical solution of the time evolution, in blue the Hamiltonian derived in Ref.~\cite{Paulisch2014}, Eq.~\eqref{eq:Hamiltonian_Paulisch}, in green the Hamiltonian derived in Ref.~\cite{Sanz2016}, Eq.~\eqref{eq:Hamiltonain_Sanz} and in red our Hamiltonian, Eq.~\eqref{eq:Hamiltonian_Sam}.
    (d) shows the relative error to the numerical solution. The exact parameters and the Hamiltonian can be found in Appendix \ref{sec:five_level} together with the formula for the relative error Eq.~\ref{eq:relative_error}.}
    \label{fig:line_plot}
\end{figure}

Fig.~\ref{fig:line_plot} shows a comparison between the three Hamiltonians, Eq.~\eqref{eq:Hamiltonian_Paulisch}, Eq.~\eqref{eq:Hamiltonain_Sanz}, and Eq.~\eqref{eq:Hamiltonian_Sam} for a five-level system with three relevant states in the limit of no time dependency. Note, that we compare our lowest order result to the second order result of Ref.~\cite{Paulisch2014} and Ref.~\cite{Sanz2016}. Our result closely follows the numerical solution, but differs slightly from the results of the others, as would be expected since the resulting Hamiltonians of the three approaches are slightly different. A more detailed discussion of the simulation can be found in Appendix \ref{sec:five_level}.

Altogether, the comparison shows that our approximation is different from the usual approaches and can not be interpreted as a straight forward generalization. However, our approach is additionally suited to deal with systems of arbitrary size and non-commuting elements as well as explicit time dependency, whereas at least one of each of these features was not covered in previous work.

 \subsection{The limit of commutativity}
As a further comparison, we show in the following that in the limit of commuting elements of the Hamiltonian our formalism reproduces the result of Ref.~\cite{Torosov2009} for the time-evolution of a two-level system.

When we move to the interaction picture with respect to the diagonal terms of the Hamiltonian Eq.~\eqref{eq:def:hamiltonian}, as shown in Appendix~\ref{appendix:interaction_picture} of the Hamiltonian, Eq.~\eqref{eq:def:hamiltonian}, the equation pair Eq.~\eqref{eq:schroedinger:equation:alpha} and Eq.~\eqref{eq:schroedinger:equation:beta} take the form
\begin{align}
         \label{eq:schroedinger:equation:alpha_prime}
        \ii\frac{\dd}{\dd t} \ket{\alpha}^{\prime} &= U_{\Delta}^{\dagger}(t_0,t)\Omega^{\dagger}U_{\Xi}(t_0,t) \ket{\beta}^{\prime}\\
        \label{eq:schroedinger:equation:beta_prime}
        \ii\frac{\dd}{\dd t} \ket{\beta}^{\prime} &= U_{\Xi}^{\dagger}(t_0,t)\Omega\, U_{\Delta}(t_0,t) \ket{\alpha}^{\prime}.
\end{align}
Following Ref.~\cite{Torosov2009}, but without the assumption of \C{non-}commutativity, we find the formal integral solution to Eq.~\eqref{eq:schroedinger:equation:beta_prime} 
\begin{align}
    \ket{\beta(t)}^{\prime} = -\ii \int\limits_{t_0}^{t}\dd s U_{\Xi}^{\dagger}(t_0,s)\Omega(s) \, U_{\Delta}(t_0,s) \ket{\alpha(s)}^{\prime}.
\end{align}
This is the formula for the relevant state found in Ref.~\cite{Torosov2009} extended to non-commuting $\Xi$, $\Delta$ and $\Omega$.

Without the assumption of non-commutativity, integration by parts as shown in Ref.~\cite{Torosov2009} is no longer possible. The lowest order solution can however be found by applying the Markov approximation $\ket{\alpha(s)}^{\prime}\approx \ket{\alpha(t)}^{\prime}$, which yields
\begin{align}
    \ket{\beta(t)}^{\prime}  = -\ii \int\limits_{t_0}^{t}\dd s U_{\Xi}^{\dagger}(t_0,s)\Omega(s) \, U_{\Delta}(t_0,s) \ket{\alpha(t)}^{\prime}.
\end{align}

When we revert the interaction picture we arrive at 
\begin{align}
        \ket{\beta(t)} = -\ii \int\limits_{t_0}^{t}\dd s U_{\Xi}(s,t)\Omega(s) \,U_{\Delta}^{\dagger}(s,t) \ket{\alpha(t)}
\end{align}
which is exactly what our first order projector predicts.
Therefore, the lowest order of our approach yields the same result as the approach in Ref.~\cite{Torosov2009} for a two-level system in the limit of commuting operators.

\section{Application}\label{sec:application}

In this section, we present our results for the first-order effective Hamiltonian for Raman diffraction.
Other typical setups, like double Raman diffraction, as well as single and double Bragg diffraction are shown in Appendix \ref{appendix:further_examples}.
In addition, we show simulation results for sine-squared pulses that show features not yet discussed or calculated in the literature.
In all those cases we find that either our formalism reproduces the results found in the literature or find corrections to the effective Hamiltonian due to the rigorous inclusion of COM degrees of freedom in our calculation. 
In all of our examples the total Hamiltonian is given by an outer product of COM degrees of freedom and internal states. 
We will represent the internal states by the typical vector notation where each dimension accounts for one state. 
However, our calculation is independent of any representation.
The choice of the vector notation is only due to the sub division of the total state and is arbitrary.
For the COM degrees of freedom we refrain from choosing any representation.

\subsection{Raman Diffraction}\label{subsec:raman}
An interesting system, not only from a technical point of view but also  from a physicists perspective is the driven  $\Lambda$-system with COM degrees of freedom. The additional COM degrees of freedom create a technical challenge as  $\Delta$ and $\Xi$ in Eq.~\eqref{eq:P:1} do not commute anymore.
This will give rise to additional recoil effects of the ancilla state, that have not yet been included into effective Hamiltonians \cite{Brion2007}.

Such setups have been featured in multiple experiments and have proved to be a successful tool in quantum optics \cite{Kasevich1991,Jones2019}.

\begin{figure}
    \centering
    \includegraphics[width=0.9\linewidth]{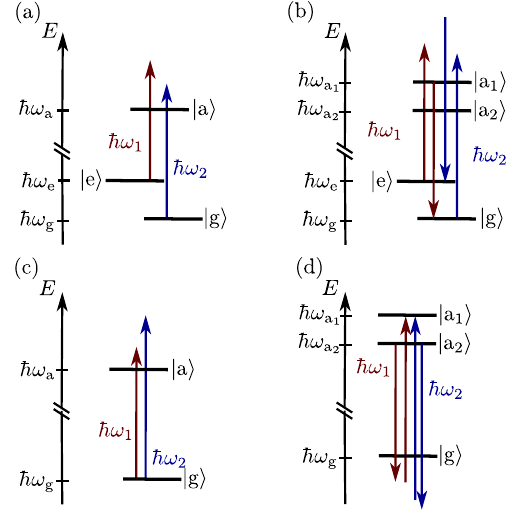}
    \caption{(a) Level scheme of a $\Lambda$-system with two lasers for Raman diffraction (b) Level scheme of a four-level system with two counter propagating lasers for Double Raman diffraction. (c) Level scheme of a two-level system with two lasers for Bragg diffraction. (d) Inverted $\Lambda$-system with four counter propagating lasers for Double Bragg diffraction.
    Note, that all systems are far detuned form the transition of the ground state manifold to the ancillas.} 
    \label{fig:level_scheme_collected}
\end{figure}

The Hamiltonian of a $\Lambda$-system coupled to a classical pair of lasers that are assumed to be plane waves in rotating wave approximation and assuming dipole interaction takes the form \cite{Moler1992, Brion2007}
\begin{align}
    H = \hbar\begin{pmatrix}
        \frac{\hat{p}^2}{2m\hbar}+ \omega_{\text{e}}&0& \Omega^{*}_{\text{e}}\ee^{-\ii k_1\hat{x}+\ii\omega_1t}\\
        0&\frac{\hat{p}^2}{2m\hbar}+ \omega_{\text{g}}& \Omega^{*}_{\text{g}}\ee^{-\ii k_2\hat{x}+\ii\omega_2t}\\
         \Omega_{\text{e}}\ee^{\ii k_1\hat{x}-\ii\omega_1t}& \Omega_{\text{g}}\ee^{\ii k_2\hat{x}-\ii\omega_2t}&\frac{\hat{p}^2}{2m\hbar}+ \omega_{\text{a}}
    \end{pmatrix}.
\end{align}
The dipole matrix elements that describe the coupling of states via the lasers are given by $\Omega_{j} \equiv \Omega_{j}(t)$. Here we assume that the laser intensity is time-dependent in form of a pulse-shape function that controls how the laser is switched on and off.
The operators $\hat{x}$ and $\hat{p}$ are the canonical COM position and momentum operator.
The system is coupled via two lasers with the wave number $k_j$ and the frequency $\omega_j$  that we indicate with the subscript 1 and 2.
Note, that each laser only couples to one of the ground states which can be achieved by choosing polarizations \cite{Hartmann2020}. The level scheme of the $\Lambda$-system including the two lasers can be seen in Fig.~\ref{fig:level_scheme_collected}.

Our first step is to move to an interaction picture via
\begin{align}
    U &= \ee^{\ii \omega_1 t} \ketbra{\text{e}}{\text{e}} + \ee^{\ii \omega_2 t} \ketbra{\text{g}}{\text{g}} + \ketbra{\text{a}}{\text{a}}\\
    \hat{H}^{\prime} &= U^{\dagger}HU-\ii\hbar U^{\dagger}\frac{\dd}{\dd t} U.
\end{align}
The resulting interaction picture Hamiltonian takes the form
\begin{align}
       H^{\prime} =\hbar \begin{pmatrix}
        \frac{\hat{p}^2}{2m \hbar}+\omega_{\text{e}} + \omega_1&0& \Omega_{\text{e}}^{*}(t)\ee^{-\ii k_1\hat{x}}\\
        0&\frac{\hat{p}^2}{2m\hbar}+\omega_{\text{g}} + \omega_2& \Omega_{\text{g}}^{*}(t)\ee^{-\ii k_2\hat{x}}\\
        \Omega_{\text{e}}(t)\ee^{\ii k_1\hat{x}}& \Omega_{\text{g}}(t)\ee^{\ii k_2\hat{x}}&\frac{\hat{p}^2}{2m \hbar}+ \omega_{\text{a}}
    \end{pmatrix}.
\end{align}
In our example at hand we are interested in the description of two-photon processes where the transition to the ancilla state is far detuned. 
We assume that the initial population is only in the excited state $\ket{\text{e}}$ or in the ground state $\ket{\text{g}}$.

The corresponding quantum state can be sub divided as 
\begin{align}
    \ket{\psi} =  \begin{pmatrix}
        \ket{\alpha}\\
        \ket{\beta}
    \end{pmatrix},\quad \text{with}\quad \ket{\alpha} = \begin{pmatrix}
        \ket{\text{e}}\\
        \ket{\text{g}}
    \end{pmatrix}\quad \text{and} \quad \ket{\beta} = \ket{\text{a}}.
\end{align}
$\ket{\alpha}$ forms the relevant subsystem containing the ground and excited state. 
The ancillate state $\ket{\text{a}}$ is treated as the irrelevant state.

Now we are in the position to interpret the operators $\Delta$, $\Xi$ and $\Omega$ according to Eq.~\eqref{eq:def:hamiltonian} as
\begin{align}
\begin{split}
    \Delta &= 
    \begin{pmatrix}
        \frac{\hat{p}^2}{2 m \hbar}+ \omega_{\text{e}}+\omega_1&0\\
        0&\frac{\hat{p}^2}{2 m \hbar}+ \omega_{\text{g}}+\omega_2
    \end{pmatrix}\\
    \Xi &= \frac{\hat{p}^2}{2 m \hbar}+ \omega_{\text{a}}\\
    \Omega &=
    \begin{pmatrix}
         \Omega_{\text{e}}\ee^{\ii k_1\hat{x}},& \Omega_{\text{g}}\ee^{\ii k_2\hat{x}}
    \end{pmatrix}
    \end{split}
\end{align}
Note that the dimensions between the operators differ and additionally the elements of 
$\Delta, \Xi$ and $\Omega$ do not commute.

In advance we define the integral
\begin{align}
    \label{eq:definition_S}
    \hat{\mathcal{S}}_{nj}(\hat{\gamma},t) = -\ii\Omega^{*}_n(t)\int\limits_{t_0}^{t} \dd s \,\Omega_j(s)\ee^{-\ii \hat{\gamma} (t-s)}.
\end{align}
Note, that this integral is central to and appears in any elimination we perform in this article. 
In fact the form of Eq.~\eqref{eq:projector_solution} generates such types of integrals \cite{Paulisch2014}. 
For some pulse shapes $\Omega_j(t)$ we are able to derive an analytic solution for this integral, as shown in Appendix~\ref{appendix:pulse_shapes}. In general no solution in closed form is available.

However, we also present methods to expand this integral under certain conditions in Appendix $\ref{appendix:pulse_shapes}$. 
In fact due to the structure of Eq.~\eqref{eq:P:1} such integrals appear for all first-order Hamiltonians \cite{Paulisch2014,Torosov2009}.

After some minor rearrangements and by using the shift property from Appendix~\ref{sec:operator_algebra} we merge the exponents in Eq.~\eqref{eq:projector_solution} into one by using the Baker-Campbell-Hausdorff formula.
We arrive at the first-order Hamiltonian 
\begin{align}
\begin{split}
    H^{\prime}_{\alpha,1} = \hbar\hat{\Delta}  +\hbar \begin{pmatrix}
        \hat{\mathcal{S}}_{\text{e}\text{e}}(\hat{\gamma}_1,t)&\ee^{\ii k\hat{x}}\hat{\mathcal{S}}_{\text{e}\text{g}}(\hat{\gamma}_2,t)\\
        \ee^{-\ii k\hat{x}}\hat{\mathcal{S}}_{\text{g}\text{e}}(\hat{\gamma}_1,t)&\hat{\mathcal{S}}_{\text{g}\text{g}}(\hat{\gamma}_2,t)
    \end{pmatrix}.
    \end{split}\label{eq:raman_result}
\end{align}
The two detunings for the diffraction processes are given by
${\hat{\gamma}_1 = \omega_{\text{a}}-\omega_{\text{e}}-\omega_1 + \hat{\nu}_1+\omega_{\text{r},1} }$ and 
${\hat{\gamma}_2 =  \omega_{\text{a}} - \omega_{\text{g}} - \omega_2 + \hat{\nu}_2 + \omega_{\text{r},2} }$, 
where ${k = k_2-k_1}$ the effective wave vector, 
${\hat{\nu}_j = \hat{p}k_j/m}$ the Doppler frequency of the laser labeled by $j$ and ${\omega_{\text{r},j} =\hbar k_j^2/(2m)}$ the respective recoil frequency.

These two detunings describe the detuning of a single photon transition to the ancilla state from either the ground or excited state. Both of these frequencies are in the optical range as we assume that we are far detuned from the transition to the ancilla state. If the detunings match the two-photon process becomes resonant.
This resonance condition differs from the usual one that was derived in the literature numerous times \cite{Moler1992,Bohringer2024} as we explicitly included the COM motion of the ancilla state.

Our calculation shows, that the two-photon light shifts $\hat{\mathcal{S}}_{\text{g}\text{g}}(\hat{\gamma}_1,t)$ and $\hat{\mathcal{S}}_{\text{e}\text{e}}(\hat{\gamma}_2,t)$ are momentum-dependent. This is due to the Doppler effect and recoil that appears when the photon is absorbed in the ancilla level or emitted from it. 
In addition, the effective two-photon Rabi frequencies $\hat{\mathcal{S}}_{ge}(\hat{\gamma}_1,t)$ and $\hat{\mathcal{S}}_{eg}(\hat{\gamma}_2,t)$ in the Hamiltonian Eq.~\eqref{eq:raman_result} are momentum-dependent. 
\par
In this case, these momentum-dependencies stem from different Doppler detunings for each absorption due to the momentum kicks imparted on the atom by the light-matter interaction. When one photon is absorbed, the momentum of the atom changes immediately. This leads to an additional Doppler detuning for the second photon. Since the two photons can be distinguished by their frequency, the order of the process plays a role, meaning that the Doppler detuning resulting from the first absorption is different from the one resulting from the second absorption. Therefore the order of the absorption has an impact on the perceived effective two photon frequency, which furthermore explains the different Doppler and recoil frequencies in the detunings $\hat{\gamma}_1$ and $\hat{\gamma}_2$.
In a typical regime for two photon transitions these Doppler shifts are rather small as the recoil and Doppler frequencies usually remain in the kHz range while the detuning to the ancilla level often is in the optical, or THz, range. Therefore, the shift from the recoil is expected to be in the order of $||\hat{\nu}_{1/2}-\omega_{k,1/2}||/(\omega_{\text{a}}-\omega_{\text{e}/\text{g}}-\omega_{1/2})\sim10^{-10}$.
\par
For a specific temporal pulse shape we are left with the task to evaluate the integral $\hat{\mathcal{S}}_{nj}(\hat{\gamma}_{\ell},t)$ for the given shape function $\Omega_j(t)$.
Note, that the operator $\hat{\gamma}_{\ell}$ depends only on the momentum operator $\hat{p}$, therefore we can disregard the operator nature in the calculation as we perform the calculation in the eigenspace of $\hat{p}$ and afterwards replace the eigenvalue with the corresponding operator again.
Noteworthy is that even for box pulses an analytical solution of the time evolution under the adiabatic Hamiltonian might not be straight forward anymore, as the two-photon Rabi frequency does not commute with the position operator. However, an analytical solution might be achievable when one follows the formalism presented in Ref.~\cite{Bohringer2024,Boehringer2025apr}.
In Appendix~\ref{appendix:pulse_shapes} we calculate  $\hat{\mathcal{S}}_{nj}(\hat{\gamma}_{\ell},t)$ for different pulse shapes. In particular we derive the analytic results for box, sine squared and Blackman pulses.

\subsection{Simulation of Time-Dependent Sine Squared Pulses}\label{subsec:sim}
A realistic choice for the temporal shape function $\Omega_j(t)$ is a sine squared pulse \cite{Shore2011} given by
\begin{align}
   \Omega_j(t) = a_0(T)\sin^2\left(\frac{\pi}{T}t\right)\quad \text{for }t\in (0,T).\label{eq:pulse_shape}
\end{align}
For such a pulse an analytic solution for the integral $\hat{\mathcal{S}}_{nj}(\hat{\gamma}_{\ell},t)$ is shown in Appendix~\ref{appendix:pulse_shapes}.
The pulse time for a two-photon Raman transition is usually defined as \cite{Bergmann1998, Vitanov2001}
\begin{align}
    A = \int\limits_{t_0}^{T}\dd t^{\prime}\, \frac{2\Omega_1(t^{\prime})\Omega_2(t^{\prime})}{\gamma_0}
\end{align}
where $\gamma_0$ is the average detuning to the ancilla state for the resonant process. In our case this is given by $\gamma_0 = (\gamma_1+\gamma_2)/2$ for the momentum eigenvalue that we want to be resonant. In our case we choose $\langle\hat{p}\rangle=0$.
We simulate the time evolution generated by the Hamiltonian Eq.~\eqref{eq:raman_result} for such a Raman setup with sine squared pulses for the $\text{D}_2$ line of ${}^{87}\text{Rb}$.

\begin{figure}
    \centering
    \includegraphics[width=
    \columnwidth]{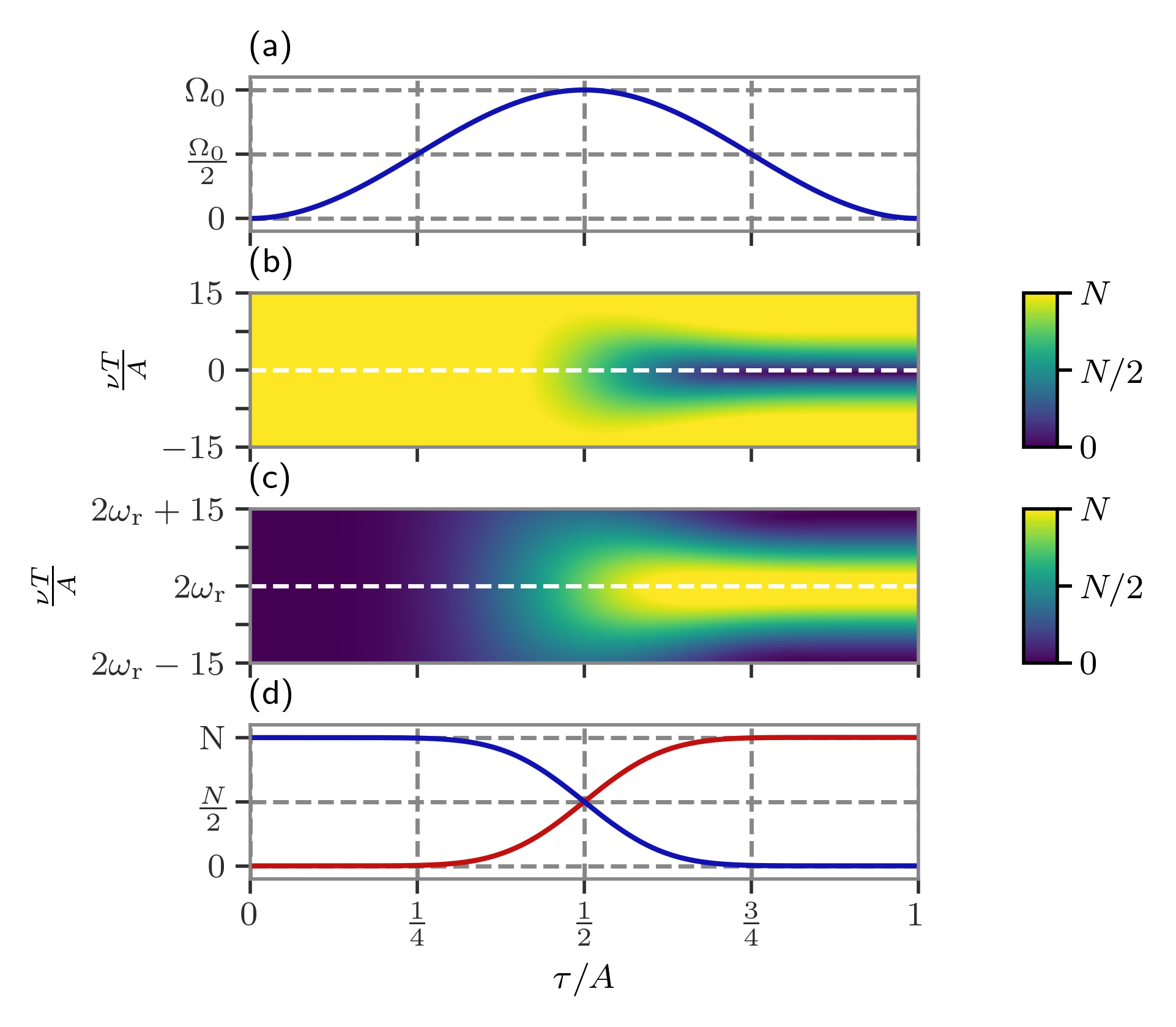}
    \caption{Raman $\pi$-pulse for the $\text{D}_2$-Line of ${}^{87}\text{Rb}$. a) shows the pulse shape of the laser in form of a sine squared pulse. b) and c) show the momentum density plot of a box wave function of the ground state in b) and the momentum shifted excited state in c). Additionally we plotted the evolution of the resonant momentum in d). These momenta are marked with the white dashed line in the density plot. For the $\pi$-pulse where the total pulse area is given by $\pi$ the population is fully inverted as long as the momentum is resonant. In the density plot the effect of velocity selectivity can clearly be seen. The parts of the wave function that are Doppler detuned from the transition oscillate with another frequency and we do not get a full population inversion. 
    The horizontal axis is the effective pulse area that was covered by the pulse so far and ends here with $A = \pi$. The vertical axis shows the Doppler frequency $\nu$ in units of the effective Rabi frequency.}
    \label{fig:plot_final_pi}
\end{figure}

\begin{figure}
    \centering
    \includegraphics[width=
    \columnwidth]{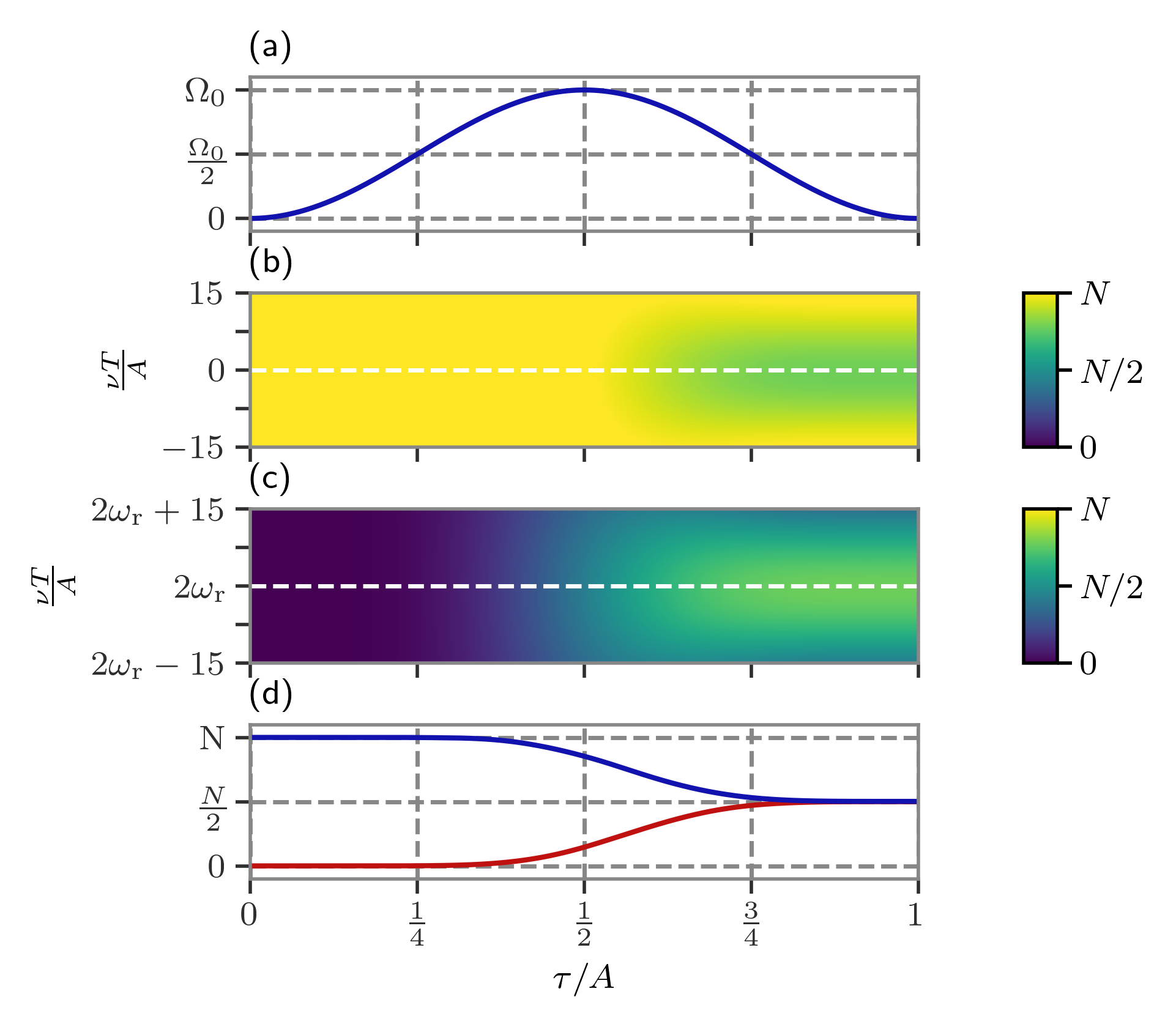}
    \caption{Raman $\pi/2$-pulse for the $\text{D}_2$-Line of ${}^{87}\text{Rb}$. a) shows the pulse shape of the laser in form of a sine squared pulse. b) and c) show the momentum density plot of a box wave function of the ground state in b) and the momentum shifted excited state in c). Additionally we plotted the evolution of the resonant momentum in d). These momenta are marked with the white dashed line in the density plot. For the $\pi/2$-pulse where the total pulse area is given by $\pi/2$ the population is brought into an equal superposition for the resonant momenta. In the density plot the effect of velocity selectivity can be observed. The horizontal axis is the effective pulse area that was covered by the pulse so far and ends here with $A = \pi/2$. The vertical axis shows the Doppler frequency $\nu$ in units of the effective Rabi frequency.}
    \label{fig:plot_final_pi2}
\end{figure}

As the Hamiltonian is explicitly time dependent we use the Runge-Kutta method of order 5(4).
For the Raman system we used the values for the $\text{D}_2$ Line of ${}^{87}\text{Rb}$ found in Ref.~\cite{steck2000sodium}.
Additionally we set the lasers to $\omega_1 = \omega_2 + \delta\omega = 247.9826227285679\, \text{THz}$ with the detuning $\delta\omega = 42943.58736018467\, \text{MHz}$.
With these values the recoil frequency $\omega_{\text{r}} = 10^{4} \,\text{Hz}$ is achieved \cite{steck2000sodium}.
As initial state, we assumed a box wave function on the total momentum grid which makes the effects of velocity selectivity more visible in plots.
The results for a  $\pi/2$-pulse and a  $\pi$-pulse can be seen in Fig.~\ref{fig:plot_final_pi2} and Fig.~\ref{fig:plot_final_pi}.
Additionally, we show the evolution of the population of the resonant momenta as line plots in Figs.~3(d) and 4(d).
Both plots show the evolution of the states under the time-dependent pulse shapes.

The more interesting feature of the effective Hamiltonian we derived is, how the Rabi frequency and the light shifts depend on the momentum or the detuning itself. As the Doppler frequency adds to the overall detuning in Eq.~\eqref{eq:definition_S} we can interpret the detuning dependency of $\mathcal{S}(\gamma,t)$ as an indirect dependency on the momentum operator. This leads to momentum-dependent lightshifts and Rabi frequencies in the effective Hamiltonian Eq.~\eqref{eq:raman_result}.

Note, that this effect comes from taking the kinetic energy of the ancilla state into account within the adiabatic elimination.
The full analytic solution of the effective evolution remains an open question that we have to leave for further investigation and publications.

\section{Conclusion and Outlook}
In section~\ref{sec:projector:method} we showed how the projector formalism developed by Sanz et. al in Ref.~\cite{Sanz2016} can be modified for temporal pulse shapes. 
This modified approach leads to a series expansion of a time dependent projector, allowing us to derive an effective Hamiltonian for the relevant sub system only. In section~\ref{sec:connection}, we showed that this formalism agrees with previous approaches when applied to the more limited cases in which they are valid.
In section~\ref{sec:application} we applied this formalism to a typical set up in matter-wave optics, i.e. Raman diffraction. 
We additionally derived the effective Hamiltonians for double Raman diffraction as well as single and double Bragg diffraction in Appendix~\ref{appendix:further_examples}. 
In particular we derived the explicit time and momentum dependency of the two-photon light shifts and Rabi frequencies within the effective Hamiltonian. 
These results go beyond the typical derivation where not only the kinetic energy of the ancilla state is neglected or treated numerically \cite{Hartmann2020,Moler1992}.
\par
Overall the method we provide has a large range of potential applications in many quantum optics systems where time-dependent pulses are of the essence or where the COM degrees of freedom can not be neglected. Especially since the dimensions of the system are not relevant to our approach as long as the conditions for the elimination are met.

\begin{acknowledgments}
The authors thank W. P. Schleich for his stimulating input and continuing support. We also thank A. Friedrich, E. Giese, G. Janson, B. Pacolli, F. Di Pumpo, and M. Schäuble for many valuable discussions. The QUANTUS and INTENTAS projects are supported by the German Space Agency at the German Aerospace Center (Deutsche Raumfahrtagentur im Deutschen Zentrum für Luft- und Raumfahrt, DLR) with funds provided by the Federal Ministry for Economic Affairs and Climate Action (Bundesministerium für Wirtschaft und Klimaschutz, BMWK) due to an enactment of the German Bundestag under Grants No. 50WM2250D (QUANTUS+) and No. 50WM2450D (QUANTUS-VI), as well as No. 50WM2178 (INTENTAS).

\section*{Data Availablility}
The data that support the findings of this study are available from the corresponding author
upon reasonable request.

\section*{Author Contributions}

S.B. was responsible for conceptualization (equal), formal analysis (lead), investigation (lead), methodology (lead), validation (equal), visualization (lead), writing—original draft (lead), and writing—review and editing (equal). A.B. was responsible for methodology (supporting), validation (supporting), and writing—review and editing (supporting). E.P.G. was responsible for conceptualization (equal), investigation (equal), methodology (supporting), validation (equal), supervision (lead), writing—original draft (supporting), and writing—review and editing (equal).

The authors have no conflicts to disclose.

\end{acknowledgments}

\appendix
\onecolumngrid

\newpage
\section{Conditions for validity of the Adiabatic Approximation}\label{sec:proof}
The conditions for a adiabatic approximation for time-independent manifolds can be derived by using a Riemann-Lebesgue type argument.
The perturbative series is generated by integrals of the form
\begin{align}
       P_{\ell}(t) = -\ii \int\limits_{t_0}^{t}\dd s U_{\Xi}(s,t)\mathcal{F}_{\ell}(s)U_{\Delta}^{\dagger}(s,t)
\end{align}
Let us for now assume that $\Xi$ and $\Delta$ are time-independent. In the time-independent case we use the decomposition of the unitary time evolution operators
\begin{align}
    U_{\Xi}(s,t) = \int\limits_{\sigma(\Xi)}\dd \mu_{\Xi}(\xi) \ketbra{\xi}{\xi}\ee^{-\ii (t-s)\xi } \\
      U_{\Delta}(s,t) = \int\limits_{\sigma(\Delta)}\dd \mu_{\Delta}(\delta) \ketbra{\delta}{\delta}\ee^{-\ii (t-s)\delta }.
    \end{align}
Let us now define two quantities, which we call the manifold gap and the intra-manifold width as 
\begin{align}
    \gamma_{\star} = \min\limits_{\sigma(\Delta),\sigma(\Xi)} (\xi - \delta)\\
    \epsilon(\xi,\delta) = \xi - \delta-\gamma_{\star}.
\end{align}
Without loss of generality we assume $\gamma_{\star}>0$.
Here $\gamma_{\star}$ describes the minimal gap between the spectra $\sigma(\Xi)$ and $\sigma(\Delta)$, while $\epsilon(\xi,\delta)$ accounts for the distance of the individual levels in the spectra to this minimal difference. 

With these definitions we find
\begin{align}
    P_{\ell}(t) = -\ii \int\limits_{\sigma(\Xi)}\dd \mu_{\Xi}(\xi) \int\limits_{\sigma(\Delta)}\dd \mu_{\Delta}(\delta) \int\limits_{t_0}^{t}\dd s\, \ee^{-\ii (t-s)\gamma_{\star}}  \bra{\xi}\mathcal{F}_{\ell}(s) \ee^{-\ii (t-s) \epsilon(\xi,\delta)}\ket{\delta} \ketbra{\xi}{\delta}\label{eq:spectral}
\end{align}
In the following we consider the kernel of the spectral integrals. That is the time integration. For a shortened notation we also introduce the quantity
\begin{align}
    F_{\ell}(\xi,\delta,s,t) = \bra{\xi}\mathcal{F}_{\ell}(s) \ee^{-\ii (t-s) \epsilon(\xi,\delta)}\ket{\delta}\ketbra{\xi}{\delta}. 
\end{align}
With this the projector $P_{\ell}(t)$ takes the form
\begin{align}
    P_{\ell}(t) = -\ii \int\limits_{\sigma(\Xi)}\dd \mu_{\Xi}(\xi) \int\limits_{\sigma(\Delta)}\dd \mu_{\Delta}(\delta)\, \int\limits_{t_0}^{t}\dd s\, \ee^{-\ii (t-s)\gamma_{\star}} F_{\ell}(\xi,\delta,s,t)  .
\end{align}
Now we can use the Riemann-Lebesgue argument to see how this integral behaves.
The substitution $s \rightarrow s^{\prime}=s+\pi / \gamma_{\star}$ yields.

\begin{align}
\begin{split}
    P_{\ell}(t)=&\ii\int\limits_{\sigma(\Xi)}\dd \mu_{\Xi}(\xi) \int\limits_{\sigma(\Delta)}\dd \mu_{\Delta}(\delta)\,\int\limits_{t_0+\pi/\gamma_{\star}}^{t+\pi/\gamma_{\star}}\dd s^{\prime}\, \ee^{-\ii (t-s^{\prime})\gamma_{\star}} F_{\ell}(\xi,\delta,s^{\prime}-\pi/\gamma_{\star},t)\\
    =&\ii\int\limits_{\sigma(\Xi)}\dd \mu_{\Xi}(\xi) \int\limits_{\sigma(\Delta)}\dd \mu_{\Delta}(\delta)\,\int\limits_{t_0}^{t}\dd s^{\prime}\, \ee^{-\ii (t-s^{\prime})\gamma_{\star}} F_{\ell}(\xi,\delta,s^{\prime}-\pi/\gamma_{\star},t) \\
    &+\ii\int\limits_{\sigma(\Xi)}\dd \mu_{\Xi}(\xi) \int\limits_{\sigma(\Delta)}\dd \mu_{\Delta}(\delta)\,\int\limits_{t}^{t+\pi/\gamma_{\star}}\dd s^{\prime}\, \ee^{-\ii (t-s^{\prime})\gamma_{\star}} F_{\ell}(\xi,\delta,s^{\prime}-\pi/\gamma_{\star},t) \\
    & -\ii\int\limits_{\sigma(\Xi)}\dd \mu_{\Xi}(\xi) \int\limits_{\sigma(\Delta)}\dd \mu_{\Delta}(\delta)\,\int\limits_{t_0}^{t_0+\pi/\gamma_{\star}}\dd s^{\prime}\, \ee^{-\ii (t-s^{\prime})\gamma_{\star}} F_{\ell}(\xi,\delta,s^{\prime}-\pi/\gamma_{\star},t) 
    \end{split}
\end{align}
Together with its original form we find
\begin{align}
\begin{split}
   2P_{\ell}(\xi,\delta,t)=& -\ii \int\limits_{\sigma(\Xi)}\dd \mu_{\Xi}(\xi) \int\limits_{\sigma(\Delta)}\dd \mu_{\Delta}(\delta)\, \int\limits_{t_0}^{t}\dd s\, \ee^{-\ii (t-s)\gamma_{\star}} [F_{\ell}(\xi,\delta,s,t) - F_{\ell}(\xi,\delta,s-\pi/\gamma_{\star},t)]\\
   &+\ii\int\limits_{\sigma(\Xi)}\dd \mu_{\Xi}(\xi) \int\limits_{\sigma(\Delta)}\dd \mu_{\Delta}(\delta)\,\int\limits_{t}^{t+\pi/\gamma_{\star}}\dd s\, \ee^{-\ii (t-s)\gamma_{\star}} F_{\ell}(\xi,\delta,s-\pi/\gamma_{\star},t) \\
    & -\ii\int\limits_{\sigma(\Xi)}\dd \mu_{\Xi}(\xi) \int\limits_{\sigma(\Delta)}\dd \mu_{\Delta}(\delta)\,\int\limits_{t_0}^{t_0+\pi/\gamma_{\star}}\dd s\, \ee^{-\ii (t-s)\gamma_{\star}} F_{\ell}(\xi,\delta,s-\pi/\gamma_{\star},t) 
   \end{split}
\end{align}
Now the last two terms can be bounded by
\begin{align}
%    \norm{-\int\limits_{t}^{t+\pi/\gamma}\dd s\, \ee^{-\ii (t-s)\gamma} F_{\ell}(\xi,\delta,s-\pi/\gamma,t)  +\int\limits_{t_0}^{t_0+\pi/\gamma}\dd s\, \ee^{-\ii (t-s)\gamma} F_{\ell}(\xi,\delta,s-\pi/\gamma,t)} \leq 2 \norm{F_{\ell}(\xi,\delta,s,t)} \pi/\gamma.
    \norm{-\int\limits_{t}^{t+\pi/\gamma_{\star}}\dd s\, ...  +\int\limits_{t_0}^{t_0+\pi/\gamma_{\star}}\dd s\, ...}_{\infty} \leq 2\frac{\pi}{\gamma_{\star}}\norm{\,\int\limits_{\sigma(\Xi)}\dd \mu_{\Xi}(\xi) \int\limits_{\sigma(\Delta)}\dd \mu_{\Delta}(\delta)\,F_{\ell}(\xi,\delta,s-\pi/(\gamma_{\star}),t)}_{\infty} .
\end{align}
The real interesting insight follows from the kernel of the first term, that is

\begin{align}
    F_{\ell}(\xi,\delta,s,t) - F_{\ell}(\xi,\delta,s-\pi/\gamma_{\star},t) = \bra{\xi}\mathcal{F}_{\ell}(s) -\mathcal{F}_{\ell}(s-\pi/\gamma_{\star})\ee^{-\ii \pi \epsilon(\xi,\delta) /\gamma_{\star}}\ket{\delta}\ketbra{\xi}{\delta}
\end{align}
which will become small if:
\begin{enumerate}[i)]
    \item $\epsilon(\xi,\delta) /\gamma_{\star}\ll1$ the difference in frequency of the level within the manifolds is small compared to the minimal gap.
    \item $\mathcal{F}(s)$ does not change considerable over a time interval of $\pi/\gamma_{\star}$. Meaning that it should be smooth on this time scale.
\end{enumerate}
This must hold for all $\xi$ and $\delta$.

With this we find 
\begin{align}
\begin{split}
    \norm{P_{\ell}(t)}\leq& \frac{\pi}{\gamma_{\star}}  \norm{\,\int\limits_{\sigma(\Xi)}\dd \mu_{\Xi}(\xi) \int\limits_{\sigma(\Delta)}\dd \mu_{\Delta}(\delta)\,\bra{\xi}\mathcal{F}_{\ell}(s) \ket{\delta}\ketbra{\xi}{\delta}}_{\infty} \\&+ \frac{(t-t_0)}{2} \norm{\,\int\limits_{\sigma(\Xi)}\dd \mu_{\Xi}(\xi) \int\limits_{\sigma(\Delta)}\dd \mu_{\Delta}(\delta)\,\bra{\xi}\mathcal{F}_{\ell}(s) -\mathcal{F}_{\ell}(s-\pi/\gamma_{\star})\ee^{-\ii \pi \epsilon(\xi,\delta) /\gamma_{\star}}\ket{\delta}\ketbra{\xi}{\delta}}_{\infty} \mustl 1 \label{eq:bound}
    \end{split}
\end{align}
Since the first order term in our series is created by 
\begin{align}
    \mathcal{F}_1(s) = \Omega(s)
\end{align}
and all higher order terms are created by sandwiches of the form $P_{\ell}\Omega^\dagger P_{n}$, we need, that the norm of $P_{1}$ is bounded by some constant $c<1$. 
From this it follows that $P_2$ is smaller than unity since it is of the order of $c^3$ and so on for the following orders. 
It is implied that $\Omega^\dagger$ and $\Omega$ have the same mathematical properties and bounds.

In terms of physics we therefore need the following conditions:
\begin{enumerate}[i)]
    \item The minimal energy gap $\gamma_{\star}$ between both manifolds needs to be larger than the driving $\Omega$.
    \item The minimal energy gap $\gamma_{\star}$ needs to be larger than the level splitting $\epsilon(\xi,\delta)$ within the respective manifolds.
    \item The time dependency of the coupling needs to be smooth and varying slowly enough such that it does not change considerable on the time scale $\pi/\gamma_{\star}$. This implies $\pi \norm{\dd \Omega(s)/\dd s}(t-t_0)/\gamma_{\star}\ll 1$.
\end{enumerate}
This must hold for all test functions such that the norm in the state space fulfills those properties. 
Note that an upper time limit for $(t-t_0)$ is also implied by Eq.~\eqref{eq:bound}. 
The smoother the time-dependent coupling the longer the times where our approximation yields good results.

A proof for time-dependent manifolds would follow the same lines but one needs to use van der Corput's lemma to estimate the oscillator integrals.
However, since we only consider time-independent manifolds in our article we will refrain from providing a more general proof or motivation.

Note, that this result is only applicable to our example if we limit ourselves to wave functions with low kinetic energy and a limited  momentum distribution width. 
Since the kinetic energy itself is not bounded, a finite energy gap between the manifolds is only guaranteed if the wave functions have these properties. 
Otherwise the Doppler frequency could lead to a vanishing gap in the spectrum.

\section{Interaction picture transformation}\label{appendix:interaction_picture}
We transform the Hamiltonian Eq.~\eqref{eq:def:hamiltonian} into the interaction picture with respect to the unitary operator
\begin{align}
    U = \begin{pmatrix}
        U_{\Delta}(t_0,t) & 0\\
        0 &  U_{\Xi}(t_0,t)
    \end{pmatrix}.
\end{align}
The resulting interaction picture Hamiltonian is given by
\begin{align}
    H^{\prime} = \begin{pmatrix}
        0 & U_{\Delta}^{\dagger}(t_0,t)\Omega^{\dagger}U_{\Xi}(t_0,t)\\
        U_{\Xi}^{\dagger}(t_0,t)\Omega U_{\Delta}(t_0,t) & 0
    \end{pmatrix},
\end{align}
which for commuting and one-dimensional $\Delta$, $\Xi$ and $\Omega$ is equivalent to the Hamiltonian treated in Ref.~\cite{Torosov2009}.

\section{Five Level System}\label{sec:five_level}
We simulated a Hamiltonian (assuming $\hbar =1$) of a five-level system with three relevant states in the limit of no time dependency to compare the results of the elimination methods from Refs.~\cite{Sanz2016, Paulisch2014} with our approach. The system consists of the three relevant states $\ket{g}$, $\ket{e}$ and $\ket{m}$, as well as two irrelevant states $\ket{a_1}$ and $\ket{a_2}$. The Hamiltonian is of the form
\begin{equation}
    H = \begin{pmatrix}
        \Delta &\Omega^{\dagger}\\
        \Omega & \Xi
    \end{pmatrix}
\end{equation}
with 
\begin{equation}
    \Delta = \text{diag}(\omega_{\text{g}},\omega_{\text{m}},\omega_{\text{e}}),\quad \Xi = \text{diag}(\omega_{\text{a}_1},\omega_{\text{a}_2}),\quad \Omega = \begin{pmatrix}
        \Omega_1&\Omega_2&\Omega_3\\
        \Omega_4&\Omega_5&\Omega_6
    \end{pmatrix}
\end{equation}
and the values in Tab.~\ref{tab:values}. 
\begin{table}[h]
    \centering
    \caption{Numerical values for the five level system in $\text{s}^{-1}$}\label{tab:values}
    \begin{tabular}{ccccccccccc}
        $\omega_{\text{g}}$ &$\omega_{\text{m}}$ &  $\omega_{\text{e}}$&$\omega_{\text{a}_1}$ &$\omega_{\text{a}_2}$ & $\Omega_1$& $\Omega_2$&$\Omega_3$ & $\Omega_4$&  $\Omega_5$&$\Omega_6$  \\ \hline
          $-4.1$ & $-4.0$ & $8.0$& $22.0$& $23.0.0$& $1.5$& $1.5$ & $1.5$& $1.0$& $1.0$& $1.0$      
    \end{tabular}
\end{table}

We estimated the relative errors of the methods against a full numerical simulation of the Hamiltonian via numerical denationalization.
The relative error of the respective method is calculated via
\begin{align}
    \delta_{\text{method}} = \left( \sum\limits_{j  ={\text{e,g,m}}} \abs{\psi_{\text{numeric}}(j,t)-\psi_{\text{method}}(j,t)}^2\right)^{\frac{1}{2}}\label{eq:relative_error}
\end{align}
where the coefficients are given by $\psi(j,t) = \braket{j}{\psi(t)}$.

\section{Operator Algebra}
\label{sec:operator_algebra}

In this section, we briefly discuss an operator identity that appears repeatedly in the preceding sections by outlining its derivation. Consider the following product of operators
\begin{align}
    \begin{split}
        \label{eq:operator:relation}
        \ee^{-\ii (t-s)(\frac{\hat{p}^2}{2m\hbar}+\omega_2)}\ee^{\ii k \hat{x}}\ee^{+\ii (t-s)(\frac{\hat{p}^2}{2m\hbar}+\omega_1)}
        &= \ee^{-\ii (t-s)\Delta\omega}\ee^{-\ii (t-s)\frac{\hat{p}^2}{2m\hbar}}\ee^{\ii k \hat{x}}\ee^{\ii (t-s)\frac{\hat{p}^2}{2m\hbar}}\\
        &= \ee^{-\ii (t-s)\Delta\omega}\ee^{\ii k (\hat{x}-(t-s)\frac{\hat{p}}{m})} \\
        &= \ee^{\ii k \hat{x}}\ee^{-\ii (t-s)(\Delta\omega+\hat{\nu}+\omega_{\text{r}})}
    \end{split}
\end{align}
where we have defined the Doppler frequency $\hat{\nu} = k\hat{p}/m$, the recoil frequency $\omega_{\text{r}} = \hbar k^2/(2m)$, and the frequency difference $\Delta\omega = \omega_2 - \omega_1$.
In the derivation of Eq.~\eqref{eq:operator:relation}, we applied the Baker–Campbell–Hausdorff (BCH) formula and Campbell’s lemma together with the canonical commutation relation $[\hat{x}, \hat{p}] = \ii \hbar$ for the COM position and momentum operator.

\section{Formulae for specific pulse shapes}\label{appendix:pulse_shapes}
In the following we briefly present the results for the central expression Eq.~\eqref{eq:definition_S} for typical windowed pulse shapes. Specifically we show the results for a box pulse, the sine squared pulse that is also discussed in Sec.~\ref{subsec:sim} and a Blackman pulse.
\subsection{Box Pulses}
For box pulses the shape function is given by 
\begin{align}
    \Omega_j(t) = a_0 \theta[t-t_0].
\end{align}
We assume that the detuning is large and in the optical frequency range, while the duration of the pulse is several orders of magnitude longer than the inverse of the detuning \cite{Kasevich1991,Ahlers2016Apr}.
Then 
\begin{align}
    \hat{\mathcal{S}}_{nj}(\hat{\gamma},t) = -\frac{a_0^2}{\hat{\gamma}}\left[1-\ee^{-\ii \hat{\gamma}(t-t_0)}\right]\approx -\frac{a_0^2}{\hat{\gamma}}
\end{align}
since the second term is oscillating on optical time scale and averages out when integrated over in time evolution during the pulse.

\subsection{Sine Squared Pulse}
The sine squared pulse is given by
\begin{align}
    \Omega_j(t)= a_0\sin^2\left(\frac{\pi (t-t_0)}{T}\right) \quad t\in(t_0,t_0+T),\quad 0 \text{ else}
\end{align}
\begin{align}
\begin{cases}
    \Omega_j(t)= a_0\sin^2\left(\frac{\pi (t-t_0)}{T}\right) \quad t\in(t_0,t_0+T),\\
     \Omega_j(t)=0 \quad\text{ else}
\end{cases}
\end{align}

where $T$ is the specific pulse parameter that indicates how fast the pulse is over. The closed form of Eq.~\eqref{eq:definition_S} is then given by
\begin{align}
\begin{split}
     \hat{\mathcal{S}}_{nj}(\hat{\gamma},t) =& -\frac{1}{2}a_0^2\sin^2\left(\frac{\pi (t-t_0)}{T}\right)\Big[\frac{1-\ee^{-\ii \hat{\gamma}(t-t_0)}}{\hat{\gamma}}-\frac{1}{2}\Big(\frac{\ee^{\ii 2\pi(t-t_0)/T}-\ee^{-\ii \hat{\gamma}(t-t_0)}}{\hat{\gamma}+2\pi/T}+\frac{\ee^{-\ii 2\pi(t-t_0)/T}-\ee^{-\ii \hat{\gamma}(t-t_0)}}{\hat{\gamma}-2\pi/T}\Big)
     \Big].
     \end{split}
\end{align}

For $\hat{\gamma} \gg \pi/T $ we can neglect the rapidly oscillating second terms in the spirit of a rotating wave approximation and we approximate $\hat{\gamma}\pm 2\pi/T\approx\hat{\gamma}$. Within this approximation we find
\begin{align}
    \hat{\mathcal{S}}_{nj}(\hat{\gamma},t)\approx -\frac{a_0^2}{\hat{\gamma}}\sin^4\left(\frac{\pi t}{T}\right)
\end{align}

which is what we expect in a lowest order approximation\cite{Torosov2009}. 

\subsection{Blackman Pulse}
For a Blackman pulse with the shape function 
\begin{align}
    \Omega_j(t) = a_0-a_1\cos(\frac{2\pi (t-t_0)}{T}) + a_2\cos(\frac{4\pi (t-t_0)}{T}) \quad t\in(t_0,t_0+T),\quad 0 \text{ else}
\end{align}
\begin{align}
\begin{cases}
    \Omega_j(t) = a_0-a_1\cos(\frac{2\pi (t-t_0)}{T}) + a_2\cos(\frac{4\pi (t-t_0)}{T}) \quad t\in(t_0,t_0+T),\\
    \Omega_j(t) =0 \quad \text{ else,}
\end{cases}
\end{align}
we find 
\begin{align}
    \begin{split}
        \hat{\mathcal{S}}_{nj}(\hat{\gamma},t) = -\Omega_n(t)\Big[& a_0\frac{1-\ee^{-\ii \hat{\gamma}(t-t_0)}}{\hat{\gamma}}-\frac{a_1}{2}\Big(\frac{\ee^{\ii 2\pi(t-t_0)/T}-\ee^{-\ii \hat{\gamma}(t-t_0)}}{\hat{\gamma}+2\pi/T}+\frac{\ee^{-\ii 2\pi(t-t_0)/T}-\ee^{-\ii \hat{\gamma}(t-t_0)}}{\hat{\gamma}-2\pi/T}\Big)\\
        &+\frac{a_2}{2}\Big(\frac{\ee^{\ii t4\pi/T}-\ee^{-\ii \hat{\gamma}(t-t_0)}}{\hat{\gamma}+4\pi/T}+\frac{\ee^{-\ii 4\pi(t-t_0)/T}-\ee^{-\ii \hat{\gamma}(t-t_0)}}{\hat{\gamma}-4\pi/T}\Big)\Big]
    \end{split}
\end{align}

\section{Further Applications}\label{appendix:further_examples}

In the following further examples we refrain from a detailed simulation and discussion of specific pulse shapes. 
Since most of our results rely on the form of $\mathcal{S}(\gamma,t)$, as defined in Eq.~\eqref{eq:definition_S}, it transfers to the following examples. 
Here the operator $\mathcal{S}(\gamma,t)$ will play the same crucial role.

\subsection{Double Raman diffraction}

In complete analogy to the Raman example of Sec.~\ref{subsec:raman} we treat the double Raman system with four lasers and two ancilla states in a retro-reflective geometry. 
The polarization is chosen such that the reflected light field is interacting with the other state in the relevant sub-state \cite{Hartmann2020}. 
This can be achieved using a $\lambda/4$ plate to change the polarization of the reflected beams \cite{Hartmann2020,Ahlers2016Apr,Leveque2009}.
In consequence we have four light fields that interact with a four level system where two lasers have the same frequency but opposite propagation directions.

The Hamiltonian for this system takes the form \cite{Giese2013,Giese2015,Ahlers2016Apr}
\begin{align}
    H^{\prime} = \hbar \begin{pmatrix}
        \Delta&\Omega^{\dagger}\\
        \Omega & \Xi
    \end{pmatrix}
\end{align}
with
\begin{align}
    \Delta &= \begin{pmatrix}
        \frac{\hat{p}^2}{2m\hbar}+\omega_{\text{e}}+\omega_{1}&0\\
        0 & \frac{\hat{p}^2}{2m\hbar}+\omega_{\text{g}}+\omega_{2}
    \end{pmatrix}\\
    \Xi &= \begin{pmatrix}
        \frac{\hat{p}^2}{2m\hbar}+\omega_{\text{a}_1}&0\\
        0 & \frac{\hat{p}^2}{2m\hbar}+\omega_{\text{a}_2}
    \end{pmatrix}\\
    \Omega &= \begin{pmatrix}
    \Omega_{\text{e}}\ee^{\ii k_{1} \hat{x}} & \Omega_{\text{g}}\ee^{\ii k_{2} \hat{x}} \\
    \Omega_{\text{e}}\ee^{-\ii k_{1} \hat{x}} & \Omega_{\text{g}}\ee^{-\ii k_{2} \hat{x}}
    \end{pmatrix}.
\end{align}
In this case we subdivide the quantum states as
\begin{align}
    \ket{\psi} =  \begin{pmatrix}
        \ket{\alpha}\\
        \ket{\beta}
    \end{pmatrix},\quad \text{with}\quad \ket{\alpha} = \begin{pmatrix}
        \ket{\text{e}}\\
        \ket{\text{g}}
    \end{pmatrix},\quad \ket{\beta} = \begin{pmatrix}
        \ket{\text{a}_1}\\
        \ket{\text{a}_2}
    \end{pmatrix}.
\end{align}
Now we are in the position to eliminate both ancilla states $ \ket{\text{a}_1}$ and $ \ket{\text{a}_2}$ simultaneously. 
A calculation analogous to the example for Raman diffraction from section~\ref{subsec:raman} results in the effective two-by-two Hamiltonian
\begin{align}
    H_{\alpha,1} =  \hbar \Delta +\hbar\mathcal{S}.
\end{align}
Here $\mathcal{S}$ is a two-by-two matrix that contains integrals of the form of Eq.~\eqref{eq:definition_S}. In that case we find the elements
\begin{subequations}
    \begin{align}
        \mathcal{S}_{11} &= \hat{\mathcal{S}}_{\text{ee}}(\hat{\gamma}_{11},t) +  \hat{\mathcal{S}}_{\text{ee}}(\hat{\gamma}_{21},t) \\
        \mathcal{S}_{12} &= \ee^{\ii k \hat{x}}\hat{\mathcal{S}}_{\text{eg}}(\hat{\gamma}_{12},t) +  \ee^{-\ii k \hat{x}}\hat{\mathcal{S}}_{\text{eg}}(\hat{\gamma}_{22},t)\\
        \mathcal{S}_{21} &= \ee^{-\ii k \hat{x}}\hat{\mathcal{S}}_{\text{ge}}(\hat{\gamma}_{11},t) + \ee^{\ii k \hat{x}} \hat{\mathcal{S}}_{\text{ge}}(\hat{\gamma}_{21},t) \\
        \mathcal{S}_{22} &= \hat{\mathcal{S}}_{\text{gg}}(\hat{\gamma}_{12},t) +  \hat{\mathcal{S}}_{\text{gg}}(\hat{\gamma}_{22},t).
    \end{align}
\end{subequations}

The detuning frequencies that contain the Doppler operator are given by
\begin{subequations}
    \begin{align}
        \hat{\gamma}_{11} &= 
        \hat{\nu}_{1}+\omega_{\text{a}_1}-\omega_{\text{e}}-\omega_{1} + \omega_{r,1} \\ 
        \hat{\gamma}_{12} &= \hat{\nu}_{2}+\omega_{\text{a}_1}-\omega_{\text{e}}-\omega_{1} + \omega_{r,2}\\
        \hat{\gamma}_{21} &=-\hat{\nu}_{1}+\omega_{\text{a}_2}-\omega_{\text{g}}-\omega_{2} + \omega_{r,1} \\
        \hat{\gamma}_{22} &=-\hat{\nu}_{2}+\omega_{\text{a}_2}-\omega_{\text{g}}-\omega_{2} + \omega_{r,2}
    \end{align}
\end{subequations}
where the recoil frequency $ \omega_{r,j} = \frac{\hbar k_j^2}{2m}$, and the Doppler operator $\hat{\nu}_j = \frac{k_j\hat{p}}{m}$ are defined as mentioned in Sec.\ref{subsec:raman}.
Note, that again the operator $\hat{\mathcal{S}}_{nj}(\hat{\gamma},t)$ plays a crucial role and takes exactly the same form as in the case of our previous example where we defined it in Eq.~\eqref{eq:definition_S}. 
However, in the previous case only one pair of lasers and a single ancilla state were present and an analytic solution can be found for many pulse shapes.
Again we find a modified Hamiltonian, that contains momentum-dependent light shifts and also momentum-dependent effective couplings between the ground and the excited state.

\subsection{Bragg diffraction}
With the typical Raman setups settled we turn to another tool in atom optics: Bragg diffraction \cite{Ahlers2016Apr,Giese2015,Giese2013,Hartmann2020_2}. 
So far we dealt with a $\Lambda$-scheme, however Bragg diffraction relies on a two-photon process where the internal state is not changed \cite{Giese2013,Giese2015}. 
Typically we have two lasers interacting with a two-level system. Again we eliminate the dynamics of the ancilla state and are only interested in the dynamics of the ground state.
Therefore we subdivide the quantum state $\ket{\psi}$ into
\begin{align}
    \ket{\psi} =  \begin{pmatrix}
        \ket{\alpha}\\
        \ket{\beta}
    \end{pmatrix},\quad \text{with}\quad \ket{\alpha} = \ket{\text{g}},\quad \ket{\beta} = \ket{\text{a}}.
\end{align}

In this representation the Hamiltonian for Bragg diffraction is given by \cite{Giese2013,Giese2015,Hartmann2020_2}
\begin{align}
\begin{split}
    H =& \sum\limits_{j = \{\text{a},\text{g}\}}\Big(\frac{\hat{p}^2}{2m} + \hbar \omega_{j}\Big)\otimes\ketbra{j}{j}  \\
    &+\hbar\Big( \sum\limits_{\ell  = \{1,2\}} \Omega^{*}_\ell\ee^{-\ii k_\ell\hat{x}+\ii\omega_\ell t}\otimes \ketbra{\text{g}}{\text{a}} + \text{h.c.}\Big).
    \end{split}
\end{align}

In the interaction picture with respect to 
\begin{align}
    U =  \ketbra{\text{a}}{\text{a}}\ee^{-\ii\omega_1t} + \ketbra{\text{g}}{\text{g}}
\end{align}
we find the Hamiltonian 
\begin{align}
    H^{\prime} = \hbar \begin{pmatrix}
        \Delta&\Omega^{\dagger}\\
        \Omega & \Xi
    \end{pmatrix}
\end{align}
with
\begin{align}
    \Delta &= \frac{\hat{p}^2}{2\hbar m}+\omega_{\text{g}}\\
    \Xi &= \frac{\hat{p}^2}{2\hbar m}+\omega_{\text{a}}-\omega_1\\
    \Omega &= \Omega_1\ee^{\ii k_1\hat{x}}+\Omega_2\ee^{\ii k_2\hat{x}-\ii\omega_{\text{L}} t}
\end{align}
where the difference in laser frequency is defined as $\omega_{\text{L}} = \omega_2-\omega_1$.
Here the calculation becomes even more straight forward since we lost the matrix structure entirely. 
Consequently we find the effective Hamiltonian for the ground state

\begin{align}
    \begin{split}
        H_{\alpha,1} = \hbar \Delta+\hbar \Big( &\hat{\mathcal{S}}_{11}(\hat{\gamma}_1,t)+\hat{\mathcal{S}}_{22}(\hat{\gamma}_2,t) \\
        &+ \ee^{\ii k \hat{x}}\hat{\mathcal{S}}_{12}(\hat{\gamma}_2,t)\ee^{-\ii\omega_{\text{L}}t}\\
        &+\ee^{-\ii k \hat{x}}\hat{\mathcal{S}}_{21}(\hat{\gamma}_1,t)\ee^{\ii\omega_{\text{L}}t}\Big)
    \end{split}
\end{align}
with $\Delta\omega = \omega_{\text{a}}-\omega_{\text{g}}-\omega_1$.
Again the elements $\hat{\mathcal{S}}_{nj}(\hat{\gamma},t)$ are defined as in Eq.~\eqref{eq:definition_S} and we have the two detuning frequencies
\begin{align}
    &\hat{\gamma}_2 = \hat{\nu}_2+\omega_{\text{r},2}+\Delta\omega - \omega_{\text{L}}\\
    &\hat{\gamma}_1 = \hat{\nu}_1+\omega_{\text{r},1}+\Delta\omega.
\end{align}
When we assume that both detunings are approximately the same $\hat{\gamma}_1 \approx \hat{\gamma}_2$, we can reduce our expression to the form derived in Refs.~\cite{Siemss2020, Fitzek2020} where the exponential functions are merged into a cosine function. However, our results goes beyond the typical Hamiltonian proposed in the literature and includes more COM effects.

\subsection{Double Bragg diffraction}
Another interesting experimental setup is double Bragg diffraction \cite{Ahlers2016Apr,Giese2013,Giese2015,Hartmann2020_2}.
Again we have a retro-reflective setup as in the case for double Raman diffraction and also introduce a second ancilla state \cite{Ahlers2016Apr,Hartmann2020}. 
Note that this case is very similar to the single Raman system where we also have three levels. However, we now have two ancilla states and only a single ground state that becomes the sole relevant state.
In that spirit we subdivide the total quantum state as
\begin{align}
    \ket{\psi} =  \begin{pmatrix}
        \ket{\text{g}}\\
        \ket{\beta}
    \end{pmatrix},\quad \text{with}\quad \ket{\beta} = \begin{pmatrix}
        \ket{\text{a}_1}\\
        \ket{\text{a}_2}
    \end{pmatrix}.
\end{align}
The Hamiltonian in the interaction picture with respect to 
\begin{align}
    U =  \ketbra{\text{a}_1}{\text{a}_1}\ee^{-\ii\omega_1t} + \ketbra{\text{a}_2}{\text{a}_2}\ee^{-\ii\omega_1t}+\ketbra{\text{g}}{\text{g}}
\end{align}
is given by
\begin{align}
    H^{\prime} = \hbar \begin{pmatrix}
        \Delta &\Omega^{\dagger}\\
        \Omega & \Xi
    \end{pmatrix}
\end{align}
with 
\begin{align}
    \Delta &= 
        \frac{\hat{p}^2}{2m\hbar}+\omega_{\text{g}}\\     
    \Xi &= \begin{pmatrix}
        \frac{\hat{p}^2}{2m\hbar}+\omega_{\text{a}_1}-\omega_{1}&0\\
        0 & \frac{\hat{p}^2}{2m\hbar}+\omega_{\text{a}_2}-\omega_{1}
    \end{pmatrix}\\
    \Omega &= \begin{pmatrix}
    \Omega_{1}\ee^{\ii k_{1} \hat{x}} + \Omega_{2}\ee^{\ii k_{2} \hat{x}-\ii\omega_{\text{L}}t} \\
    \Omega_{1}\ee^{-\ii k_{1} \hat{x}} + \Omega_{2}\ee^{-\ii k_{2} \hat{x}-\ii\omega_{\text{L}}t}.
    \end{pmatrix}
\end{align}
Again we defined the difference in laser frequency $\omega_{\text{L}} = \omega_{2}-\omega_1$.
The resulting effective Hamiltonian for the ground state $\ket{\text{g}}$ is given by
\begin{align}
    \begin{split}
        H_{\alpha,1} =& \hbar \Delta + \hbar \Big[ \hat{\mathcal{S}}_{11}(\hat{\gamma}_{11},t) + \hat{\mathcal{S}}_{22}(\hat{\gamma}_{12},t) \\&+ \ee^{\ii k \hat{x}-\ii\omega_{\text{L}}t}\hat{\mathcal{S}}_{12}(\hat{\gamma}_{12},t) + \ee^{-\ii k \hat{x}+\ii\omega_{\text{L}}t}\hat{\mathcal{S}}_{21}(\hat{\gamma}_{11},t)
        \\&+\hat{\mathcal{S}}_{11}(\hat{\gamma}_{21},t) + \hat{\mathcal{S}}_{22}(\hat{\gamma}_{22},t) \\&+ \ee^{-\ii k \hat{x}-\ii\omega_{\text{L}}t}\hat{\mathcal{S}}_{12}(\hat{\gamma}_{22},t) + \ee^{\ii k \hat{x}+\ii\omega_{\text{L}}t}\hat{\mathcal{S}}_{21}(\hat{\gamma}_{21},t)
        \Big]
    \end{split}
\end{align}
with the effective detuning frequencies
\begin{subequations}
    \begin{align}
        \hat{\gamma}_{11} &= 
            \hat{\nu}_{1}+\omega_{\text{a}_1}-\omega_{\text{g}}-\omega_{1} + \omega_{r,1} \\
             \hat{\gamma}_{12} &= \hat{\nu}_{2}+\omega_{\text{a}_1}-\omega_{\text{g}}-\omega_{1} +\omega_{r,2}-\omega_{\text{L}}\\
             \hat{\gamma}_{21} &=-\hat{\nu}_{1}+\omega_{\text{a}_2}-\omega_{\text{g}}-\omega_{1} + \omega_{r,1} \\
             \hat{\gamma}_{22} &=-\hat{\nu}_{2}+\omega_{\text{a}_2}-\omega_{\text{g}}-\omega_{1} +\omega_{r,2}-\omega_{\text{L}}.
    \end{align}
\end{subequations}
Again the elements $\hat{\mathcal{S}}_{nj}(\hat{\gamma},t)$ as defined in Eq.~\eqref{eq:definition_S} become the central operator in the effective Hamiltonian and we find the momentum-dependent light shifts and Rabi frequencies in complete analogy to our other examples.

\twocolumngrid

\bibliography{references}

@misc{Glasbrenner2025,
      title={Hidden facts in {L}andau-{Z}ener transitions revealed by the {R}iccati {E}quation}, 
      author={Eric P. Glasbrenner and Yannik Gerdes and Sándor Varró and Wolfgang P. Schleich},
      year={2025},
      eprint={2502.04033},
      archivePrefix={arXiv},
      primaryClass={quant-ph},
      url={https://arxiv.org/abs/2502.04033}, 
}

@article{Brion2007,
doi = {10.1088/1751-8113/40/5/011},
url = {https://dx.doi.org/10.1088/1751-8113/40/5/011},
year = {2007},
month = {jan},
publisher = {},
volume = {40},
number = {5},
pages = {1033},
author = {Brion, E and Pedersen, L H and Mølmer, K},
title = {Adiabatic elimination in a lambda system},
journal = {Journal of Physics A: Mathematical and Theoretical}
}

@article{Torosov2009,
  title = {Phase shifts in nonresonant coherent excitation},
  author = {Torosov, Boyan T. and Vitanov, Nikolay V.},
  journal = {Phys. Rev. A},
  volume = {79},
  issue = {4},
  pages = {042108},
  numpages = {9},
  year = {2009},
  month = {Apr},
  publisher = {American Physical Society},
  doi = {10.1103/PhysRevA.79.042108},
  url = {https://link.aps.org/doi/10.1103/PhysRevA.79.042108}
}

@article{Paulisch2014,
author={Paulisch, Vanessa
and Rui, Han
and Ng, Hui Khoon
and Englert, Berthold-Georg},
title={Beyond adiabatic elimination: {A} hierarchy of approximations for multi-photon processes},
journal={The European Physical Journal Plus},
year={2014},
month={Jan},
day={28},
volume={129},
number={1},
pages={12},
issn={2190-5444},
doi={10.1140/epjp/i2014-14012-8},
url={https://doi.org/10.1140/epjp/i2014-14012-8}
}

@incollection{Sanz2016,
  title={Beyond adiabatic elimination: {E}ffective {H}amiltonians and singular perturbation},
  author={Sanz, M. and Solano, E. and Egusquiza, {\'I}.},
  booktitle={Applications+ Practical Conceptualization+ Mathematics= fruitful Innovation},
  pages={127--142},
  year={2016},
  editor={Anderssen, R. S. and Broadbridge, P. and Fukumoto, Y. and Kajiwara, K. and Takagi, T. and Verbitskiy, E. and Wakayama, M.},
  publisher={Springer},
  doi = {10.1007/978-4-431-55342-7_12},
  url = {https://link.springer.com/chapter/10.1007/978-4-431-55342-7_12},
}

@article{Semin2016,
	author = {Semin, Vitalii and Petruccione, Francesco},
	title = {{Projection operator based expansion of the evolution operator}},
	journal = {J. Phys. A: Math. Theor.},
	volume = {49},
	number = {42},
	pages = {425301},
	year = {2016},
	issn = {1751-8121},
	publisher = {IOP Publishing},
	doi = {10.1088/1751-8113/49/42/425301}
}

@article{Bott2023,
    author = {Bott, Alexander and Di Pumpo, Fabio and Giese, Enno},
    title = {Atomic diffraction from single-photon transitions in gravity and Standard-Model extensions},
    journal = {AVS Quantum Science},
    volume = {5},
    number = {4},
    pages = {044402},
    year = {2023},
    month = {11},
    issn = {2639-0213},
    doi = {10.1116/5.0174258},
    url = {https://doi.org/10.1116/5.0174258}
}

@article{Bohringer2024,
	author = {B{\ifmmode\ddot{o}\else\"{o}\fi}hringer, Samuel and Friedrich, Alexander},
	title = {{Decoupling of external and internal dynamics in driven two-level systems}},
	journal = {Phys. Rev. Res.},
	volume = {6},
	number = {4},
	pages = {043153},
	year = {2024},
	publisher = {American Physical Society},
	doi = {10.1103/PhysRevResearch.6.043153}
}

@misc{Bruhnke2024,
	author = {Bruhnke, Jakob},
	title = {{Two-photon {R}abi oscillations in hydrogen: {A} theoretical study of effective {H}amiltonian approaches}},
	year = {2024},
	note = {[Online; accessed 20. May 2025]},
	url = {https://lup.lub.lu.se/student-papers/search/publication/9160248}
}

@article{Moler1992,
	author = {Moler, Kathryn and Weiss, David S. and Kasevich, Mark and Chu, Steven},
	title = {{Theoretical analysis of velocity-selective Raman transitions}},
	journal = {Phys. Rev. A},
	volume = {45},
	number = {1},
	pages = {342--348},
	year = {1992},
	publisher = {American Physical Society},
	doi = {10.1103/PhysRevA.45.342}
}

@article{Hartmann2020,
  title = {Atomic Raman scattering: Third-order diffraction in a double geometry},
  author = {Hartmann, Sabrina and Jenewein, Jens and Abend, Sven and Roura, Albert and Giese, Enno},
  journal = {Phys. Rev. A},
  volume = {102},
  issue = {6},
  pages = {063326},
  numpages = {12},
  year = {2020},
  month = {Dec},
  publisher = {American Physical Society},
  doi = {10.1103/PhysRevA.102.063326},
  url = {https://link.aps.org/doi/10.1103/PhysRevA.102.063326}
}

@article{Hartmann2020_2,
  title = {Regimes of atomic diffraction: Raman versus Bragg diffraction in retroreflective geometries},
  author = {Hartmann, Sabrina and Jenewein, Jens and Giese, Enno and Abend, Sven and Roura, Albert and Rasel, Ernst M. and Schleich, Wolfgang P.},
  journal = {Phys. Rev. A},
  volume = {101},
  issue = {5},
  pages = {053610},
  numpages = {16},
  year = {2020},
  publisher = {American Physical Society},
  doi = {10.1103/PhysRevA.101.053610},
  url = {https://link.aps.org/doi/10.1103/PhysRevA.101.053610}
}

@article{Giese2013,
  title = {Double Bragg diffraction: A tool for atom optics},
  author = {Giese, E. and Roura, A. and Tackmann, G. and Rasel, E. M. and Schleich, W. P.},
  journal = {Phys. Rev. A},
  volume = {88},
  issue = {5},
  pages = {053608},
  numpages = {23},
  year = {2013},
  publisher = {American Physical Society},
  doi = {10.1103/PhysRevA.88.053608},
  url = {https://link.aps.org/doi/10.1103/PhysRevA.88.053608}
}

@article{Giese2015,
author = {Giese, Enno},
title = {Mechanisms of matter-wave diffraction and their application to interferometers},
journal = {Fortschritte der Physik},
volume = {63},
number = {6},
pages = {337-410},
doi = {https://doi.org/10.1002/prop.201500020},
url = {https://onlinelibrary.wiley.com/doi/abs/10.1002/prop.201500020},
year = {2015}
}

@article{Siemss2020,
  title = {Analytic theory for Bragg atom interferometry based on the adiabatic theorem},
  author = {Siem\ss{}, Jan-Niclas and Fitzek, Florian and Abend, Sven and Rasel, Ernst M. and Gaaloul, Naceur and Hammerer, Klemens},
  journal = {Phys. Rev. A},
  volume = {102},
  issue = {3},
  pages = {033709},
  numpages = {26},
  year = {2020},
  publisher = {American Physical Society},
  doi = {10.1103/PhysRevA.102.033709},
  url = {https://link.aps.org/doi/10.1103/PhysRevA.102.033709}
}

@article{Fitzek2020,
  author       = {Fitzek, Florian and Siemß, Jan-Niclas and Seckmeyer, Stefan and Ahlers, Holger and Rasel, Ernst M. and Hammerer, Klemens and Gaaloul, Naceur},
  title        = {Universal atom interferometer simulation of elastic scattering processes},
  journal      = {Scientific Reports},
  volume       = {10},
  number       = {1},
  pages        = {22120},
  year         = {2020},
  doi          = {10.1038/s41598-020-78859-1},
  url          = {https://doi.org/10.1038/s41598-020-78859-1},
  issn         = {2045-2322}
}

@article{Li2024,
  title = {Robust double Bragg diffraction via detuning control},
  author = {Li, Rui and Mart\'{\i}nez-Lahuerta, V. J. and Seckmeyer, S. and Hammerer, Klemens and Gaaloul, Naceur},
  journal = {Phys. Rev. Res.},
  volume = {6},
  issue = {4},
  pages = {043236},
  numpages = {16},
  year = {2024},
  publisher = {American Physical Society},
  doi = {10.1103/PhysRevResearch.6.043236},
  url = {https://link.aps.org/doi/10.1103/PhysRevResearch.6.043236}
}

@article{fausett2009sylvester,
  title={On Sylvester matrix differential equations, Analytical and numerical solutions},
  author={Fausett, Laurene V},
  journal={International Journal of Pure and Applied Mathematics},
  doi = {}, 
  volume={53},
  number={1},
  pages={55--68},
  year={2009}
}

@article{Behr2019Dec,
	author = {Behr, Maximilian and Benner, Peter and Heiland, Jan},
	title = {{Solution formulas for differential Sylvester and Lyapunov equations}},
	journal = {Calcolo},
	volume = {56},
	number = {4},
	pages = {1--33},
	year = {2019},
	month = dec,
	issn = {1126-5434},
	publisher = {Springer International Publishing},
	doi = {10.1007/s10092-019-0348-x}
}

@article{steck2000sodium,
  title={Sodium D Line Data},
  author={Steck, Daniel A.},
  journal={available online at https://steck.us/alkalidata/rubidium87numbers.1.6.pdf},
  year={revision 2.3.3, 28 May 2024},
  url={https://steck.us/alkalidata/rubidium87numbers.1.6.pdf}
}

@article{Gentile1989Nov,
	author = {Gentile, Thomas R. and Hughey, Barbara J. and Kleppner, Daniel and Ducas, Theodore W.},
	title = {{Experimental study of one- and two-photon Rabi oscillations}},
	journal = {Phys. Rev. A},
	volume = {40},
	number = {9},
	pages = {5103--5115},
	year = {1989},
	month = nov,
	publisher = {American Physical Society},
	doi = {10.1103/PhysRevA.40.5103}
}

@article{Beterov_2024,
   title={Rabi oscillations at three-photon laser excitation of a single rubidium Rydberg atom in an optical dipole trap},
   volume={},
   ISSN={},
   url={http://dx.doi.org/10.31857/S0044451024100109},
   DOI={10.48550/arXiv.2410.01703},
   number={},
   journal={arXiv},
   publisher={},
   author={Beterov, I. I and Yakshina, E. A and Suliman, G. and Betleni, P. I and Prilutskaya, A. A and Skvortsova, D. A and Zagirov, T. R and Tret’yakov, D. B and Entin, V. M and Bezuglov, N. N and Ryabtsev, I. I},
   year={2024},
   month=dec, pages={} }

@article{HU2023170637,
title = {Improving the fringe contrast in an atomic gravimeter by optimizing the Raman laser intensity},
journal = {Optik},
volume = {276},
pages = {170637},
year = {2023},
issn = {0030-4026},
doi = {https://doi.org/10.1016/j.ijleo.2023.170637},
url = {https://www.sciencedirect.com/science/article/pii/S003040262300133X},
author = {Qing-Qing Hu and Hang Zhou and Yu-Kun Luo and Qin Luo and Wen-Jun Kuang and Fu-Bin Wan and Yao-Yu Zhong and Fu-Fang Xu},
}

@book{Scullyzubairy1997,
place={Cambridge},
title={Quantum Optics},
DOI={10.1017/CBO9780511813993},
publisher={Cambridge University Press},
author={Scully, Marlan O. and Zubairy, M. Suhail},
year={1997}}

@book{Breuer2007,
    author = {Breuer, Heinz-Peter and Petruccione, Francesco},
    title = "{The Theory of Open Quantum Systems}",
    publisher = {Oxford University Press},
    year = {2007},
    month = {01},
    isbn = {9780199213900},
    doi = {10.1093/acprof:oso/9780199213900.001.0001},
    url = {https://doi.org/10.1093/acprof:oso/9780199213900.001.0001},
}

@book{schleich2011,
  title={Quantum Optics in Phase Space},
  author={Schleich, W.P.},
  isbn={9783527635009},
  url={https://books.google.de/books?id=2jUjQPW-WXAC},
  year={2011},
  publisher={Wiley}
}

@article{Nakajima1958Dec,
	author = {Nakajima, Sadao},
	title = {{On Quantum Theory of Transport Phenomena: Steady Diffusion}},
	journal = {Prog. Theor. Phys.},
	volume = {20},
	number = {6},
	pages = {948--959},
	year = {1958},
	month = dec,
	issn = {0033-068X},
	publisher = {Oxford Academic},
	doi = {10.1143/PTP.20.948}
}

@article{Zwanzig1960Nov,
	author = {Zwanzig, Robert},
	title = {{Ensemble Method in the Theory of Irreversibility}},
	journal = {J. Chem. Phys.},
	volume = {33},
	number = {5},
	pages = {1338--1341},
	year = {1960},
	month = nov,
	issn = {0021-9606},
	publisher = {AIP Publishing},
	doi = {10.1063/1.1731409}
}

@article{Ahlers2016Apr,
	author = {Ahlers, H. and M{\ifmmode\ddot{u}\else\"{u}\fi}ntinga, H. and Wenzlawski, A. and Krutzik, M. and Tackmann, G. and Abend, S. and Gaaloul, N. and Giese, E. and Roura, A. and Kuhl, R. and L{\ifmmode\ddot{a}\else\"{a}\fi}mmerzahl, C. and Peters, A. and Windpassinger, P. and Sengstock, K. and Schleich, W. P. and Ertmer, W. and Rasel, E. M.},
	title = {{Double Bragg Interferometry}},
	journal = {Phys. Rev. Lett.},
	volume = {116},
	number = {17},
	pages = {173601},
	year = {2016},
	month = apr,
	publisher = {American Physical Society},
	doi = {10.1103/PhysRevLett.116.173601}
}

@article{Savoie2018Dec,
	author = {Savoie, D. and Altorio, M. and Fang, B. and Sidorenkov, L. A. and Geiger, R. and Landragin, A.},
	title = {{Interleaved atom interferometry for high-sensitivity inertial measurements}},
	journal = {Sci. Adv.},
	volume = {4},
	number = {12},
	year = {2018},
	month = dec,
	issn = {2375-2548},
	publisher = {American Association for the Advancement of Science},
	doi = {10.1126/sciadv.aau7948}
}

@article{Freier2016Jun,
	author = {Freier, C. and Hauth, M. and Schkolnik, V. and Leykauf, B. and Schilling, M. and Wziontek, H. and Scherneck, H.-G. and M{\ifmmode\ddot{u}\else\"{u}\fi}ller, J. and Peters, A.},
	title = {{Mobile quantum gravity sensor with unprecedented stability}},
	journal = {J. Phys. Conf. Ser.},
	volume = {723},
	number = {1},
	pages = {012050},
	year = {2016},
	month = jun,
	issn = {1742-6596},
	publisher = {IOP Publishing},
	doi = {10.1088/1742-6596/723/1/012050}
}

@article{Schlippert2014,
  title = {Quantum Test of the Universality of Free Fall},
  author = {Schlippert, D. and Hartwig, J. and Albers, H. and Richardson, L. L. and Schubert, C. and Roura, A. and Schleich, W. P. and Ertmer, W. and Rasel, E. M.},
  journal = {Phys. Rev. Lett.},
  volume = {112},
  issue = {20},
  pages = {203002},
  numpages = {5},
  year = {2014},
  month = {May},
  publisher = {American Physical Society},
  doi = {10.1103/PhysRevLett.112.203002},
  url = {https://link.aps.org/doi/10.1103/PhysRevLett.112.203002}
}

@article{Ludlow2015Jun,
	author = {Ludlow, Andrew D. and Boyd, Martin M. and Ye, Jun and Peik, E. and Schmidt, P. O.},
	title = {{Optical atomic clocks}},
	journal = {Rev. Mod. Phys.},
	volume = {87},
	number = {2},
	pages = {637--701},
	year = {2015},
	month = jun,
	publisher = {American Physical Society},
	doi = {10.1103/RevModPhys.87.637}
}

@article{Kasevich1991,
  title = {Atomic interferometry using stimulated Raman transitions},
  author = {Kasevich, Mark and Chu, Steven},
  journal = {Phys. Rev. Lett.},
  volume = {67},
  issue = {2},
  pages = {181--184},
  numpages = {0},
  year = {1991},
  month = {Jul},
  publisher = {American Physical Society},
  doi = {10.1103/PhysRevLett.67.181},
  url = {https://link.aps.org/doi/10.1103/PhysRevLett.67.181}
}

@article{Oelker2019Oct,
	author = {Oelker, E. and Hutson, R. B. and Kennedy, C. J. and Sonderhouse, L. and Bothwell, T. and Goban, A. and Kedar, D. and Sanner, C. and Robinson, J. M. and Marti, G. E. and Matei, D. G. and Legero, T. and Giunta, M. and Holzwarth, R. and Riehle, F. and Sterr, U. and Ye, J.},
	title = {{Demonstration of 4.8{\hspace{0.167em}}{\ifmmode\times\else\texttimes\fi}{\hspace{0.167em}}10{-}17 stability at 1{\hspace{0.167em}}s for two independent optical clocks}},
	journal = {Nat. Photonics},
	volume = {13},
	pages = {714--719},
	year = {2019},
	month = oct,
	issn = {1749-4893},
	publisher = {Nature Publishing Group},
	doi = {10.1038/s41566-019-0493-4}
}

@article{McGrew2019Apr,
	author = {McGrew, W. F. and Zhang, X. and Leopardi, H. and Fasano, R. J. and Nicolodi, D. and Beloy, K. and Yao, J. and Sherman, J. A. and Sch{\ifmmode\ddot{a}\else\"{a}\fi}ffer, S. A. and Savory, J. and Brown, R. C. and R{\ifmmode\ddot{o}\else\"{o}\fi}misch, S. and Oates, C. W. and Parker, T. E. and Fortier, T. M. and Ludlow, A. D.},
	title = {{Towards the optical second: verifying optical clocks at the SI limit}},
	journal = {Optica},
	volume = {6},
	number = {4},
	pages = {448--454},
	year = {2019},
	month = apr,
	issn = {2334-2536},
	publisher = {Optica Publishing Group},
	doi = {10.1364/OPTICA.6.000448}
}

@article{Werner2024,
	author = {Werner, Michael and Schwartz, Philip K. and Kirsten-Siem{\ss}, Jan-Niclas and Gaaloul, Naceur and Giulini, Domenico and Hammerer, Klemens},
	title = {{Atom interferometers in weakly curved spacetimes using Bragg diffraction and Bloch oscillations}},
	journal = {Phys. Rev. D},
	volume = {109},
	number = {2},
	pages = {022008},
	year = {2024},
	month = jan,
	publisher = {American Physical Society},
	doi = {10.1103/PhysRevD.109.022008}
}

@article{DiPumpo2023mar,
  title = {Universality-of-clock-rates test using atom interferometry with ${T}^{3}$ scaling},
  author = {Di Pumpo, Fabio and Friedrich, Alexander and Ufrecht, Christian and Giese, Enno},
  journal = {Phys. Rev. D},
  volume = {107},
  issue = {6},
  pages = {064007},
  numpages = {12},
  year = {2023},
  month = mar,
  publisher = {American Physical Society},
  doi = {10.1103/PhysRevD.107.064007},
  url = {https://link.aps.org/doi/10.1103/PhysRevD.107.064007}
}

@article{DiPumpo2021nov,
  title = {Gravitational Redshift Tests with Atomic Clocks and Atom Interferometers},
  author = {Di Pumpo, Fabio and Ufrecht, Christian and Friedrich, Alexander and Giese, Enno and Schleich, Wolfgang P. and Unruh, William G.},
  journal = {PRX Quantum},
  volume = {2},
  issue = {4},
  pages = {040333},
  numpages = {23},
  year = {2021},
  month = nov,
  publisher = {American Physical Society},
  doi = {10.1103/PRXQuantum.2.040333},
  url = {https://link.aps.org/doi/10.1103/PRXQuantum.2.040333}
}

@article{Antoine2003Apr,
	author = {Antoine, Ch and Bord{\ifmmode\acute{e}\else\'{e}\fi}, Ch J.},
	title = {{Quantum theory of atomic clocks and gravito-inertial sensors: an update}},
	journal = {J. Opt. B: Quantum Semiclassical Opt.},
	volume = {5},
	number = {2},
	pages = {S199},
	year = {2003},
	month = apr,
	issn = {1464-4266},
	publisher = {IOP Publishing},
	doi = {10.1088/1464-4266/5/2/380}
}

@article{Dalibard1992Feb,
	author = {Dalibard, Jean and Castin, Yvan and M{\o}lmer, Klaus},
	title = {{Wave-function approach to dissipative processes in quantum optics}},
	journal = {Phys. Rev. Lett.},
	volume = {68},
	number = {5},
	pages = {580--583},
	year = {1992},
	month = feb,
	publisher = {American Physical Society},
	doi = {10.1103/PhysRevLett.68.580}
}

@article{Castin1991Apr,
	author = {Castin, Y. and Dalibard, J.},
	title = {{Quantization of Atomic Motion in Optical Molasses}},
	journal = {Europhys. Lett.},
	volume = {14},
	number = {8},
	pages = {761},
	year = {1991},
	month = apr,
	issn = {0295-5075},
	publisher = {IOP Publishing},
	doi = {10.1209/0295-5075/14/8/007}
}

@article{Dalibard1985Nov,
	author = {Dalibard, J. and Cohen-Tannoudji, C.},
	title = {{Dressed-atom approach to atomic motion in laser light: the dipole force revisited}},
	journal = {J. Opt. Soc. Am. B, JOSAB},
	volume = {2},
	number = {11},
	pages = {1707--1720},
	year = {1985},
	month = nov,
	issn = {1520-8540},
	publisher = {Optica Publishing Group},
	doi = {10.1364/JOSAB.2.001707}
}

@article{Dalibard1989Nov,
	author = {Dalibard, J. and Cohen-Tannoudji, C.},
	title = {{Laser cooling below the Doppler limit by polarization gradients: simple theoretical models}},
	journal = {J. Opt. Soc. Am. B, JOSAB},
	volume = {6},
	number = {11},
	pages = {2023--2045},
	year = {1989},
	month = nov,
	issn = {1520-8540},
	publisher = {Optica Publishing Group},
	doi = {10.1364/JOSAB.6.002023}
}

@article{Aspect1986Oct,
	author = {Aspect, A. and Dalibard, J. and Heidmann, A. and Salomon, C. and Cohen-Tannoudji, C.},
	title = {{Cooling Atoms with Stimulated Emission}},
	journal = {Phys. Rev. Lett.},
	volume = {57},
	number = {14},
	pages = {1688--1691},
	year = {1986},
	month = oct,
	publisher = {American Physical Society},
	doi = {10.1103/PhysRevLett.57.1688}
}

@article{Boehringer2025apr,
  title = {Evolution of momentum-dependent observables under stochastic phase noise in Rabi oscillations},
  author = {B\"ohringer, Samuel and Kienle, Fabian and Lopp, Richard},
  journal = {Phys. Rev. Res.},
  volume = {7},
  issue = {2},
  pages = {023048},
  numpages = {11},
  year = {2025},
  month = apr,
  publisher = {American Physical Society},
  doi = {10.1103/PhysRevResearch.7.023048},
  url = {https://link.aps.org/doi/10.1103/PhysRevResearch.7.023048}
}

@article{Ufrecht2020,
	author = {Ufrecht, Christian and Giese, Enno},
	title = {{Perturbative operator approach to high-precision light-pulse atom interferometry}},
	journal = {Phys. Rev. A},
	volume = {101},
	number = {5},
	pages = {053615},
	year = {2020},
	month = may,
	publisher = {American Physical Society},
	doi = {10.1103/PhysRevA.101.053615}
}

@book{Keldysh,
	author = {Keldysh, L. V.},
	title = {{Diagram technique for nonequilibrium processes}},
	booktitle = {{SOVIET PHYSICS JETP}},
	pages = {47--55},
	year = {1965},
	month = may,
	isbn = {978-981-12-7945-4},
	publisher = {WORLD SCIENTIFIC},
	address = {Singapore},
	doi = {10.1142/9789811279461_0007}
}

@book{Stefanucci2013,
	author = {Stefanucci, Gianluca and van Leeuwen, Robert},
	title = {{Nonequilibrium Many-Body Theory of Quantum Systems: A Modern Introduction}},
	journal = {Cambridge Core},
	year = {2013},
	month = mar,
	isbn = {978-0-52176617-3},
	publisher = {Cambridge University Press},
	address = {Cambridge, England, UK},
	doi = {10.1017/CBO9781139023979}
}

@article{Jones2019,
	author = {Jones, Robin R. and Hooper, David C. and Zhang, Liwu and Wolverson, Daniel and Valev, Ventsislav K.},
	title = {{Raman Techniques: Fundamentals and Frontiers}},
	journal = {Nanoscale Res. Lett.},
	volume = {14},
	number = {1},
	pages = {1--34},
	year = {2019},
	month = dec,
	issn = {1556-276X},
	publisher = {Springer US},
	doi = {10.1186/s11671-019-3039-2}
}

@article{Leveque2009,
	author = {L{\ifmmode\acute{e}\else\'{e}\fi}v{\ifmmode\grave{e}\else\`{e}\fi}que, T. and Gauguet, A. and Michaud, F. and Pereira Dos Santos, F. and Landragin, A.},
	title = {{Enhancing the Area of a Raman Atom Interferometer Using a Versatile Double-Diffraction Technique}},
	journal = {Phys. Rev. Lett.},
	volume = {103},
	number = {8},
	pages = {080405},
	year = {2009},
	month = aug,
	publisher = {American Physical Society},
	doi = {10.1103/PhysRevLett.103.080405}
}

@book{Shore2011,
	author = {Shore, Bruce W.},
	title = {{Manipulating Quantum Structures Using Laser Pulses}},
	journal = {Cambridge Core},
	year = {2011},
	month = sep,
	isbn = {978-0-52176357-8},
	publisher = {Cambridge University Press},
	address = {Cambridge, England, UK},
	doi = {10.1017/CBO9780511675713}
}

@article{Vitanov2001,
	author = {Vitanov, Nikolay V. and Halfmann, Thomas and Shore, Bruce W. and Bergmann, Klaas},
	title = {{LASER-INDUCED}},
	journal = {Annu. Rev. Phys. Chem.},
	number = {Volume 52, 2001},
	pages = {763--809},
	year = {2001},
	month = oct,
	publisher = {Annual Reviews},
	doi = {10.1146/annurev.physchem.52.1.763}
}

@article{Bergmann1998,
	author = {Bergmann, K. and Theuer, H. and Shore, B. W.},
	title = {{Coherent population transfer among quantum states of atoms and molecules}},
	journal = {Rev. Mod. Phys.},
	volume = {70},
	number = {3},
	pages = {1003--1025},
	year = {1998},
	month = jul,
	publisher = {American Physical Society},
	doi = {10.1103/RevModPhys.70.1003}
}

@article{Heavner2014,
	author = {Heavner, Thomas P. and Donley, Elizabeth A. and Levi, Filippo and Costanzo, Giovanni and Parker, Thomas E. and Shirley, Jon H. and Ashby, Neil and Barlow, Stephan and Jefferts, S. R.},
	title = {{First accuracy evaluation of NIST-F2}},
	journal = {Metrologia},
	volume = {51},
	number = {3},
	pages = {174},
	year = {2014},
	month = may,
	issn = {0026-1394},
	publisher = {IOP Publishing},
	doi = {10.1088/0026-1394/51/3/174}
}

@article{Abend2024,
	author = {Abend, Sven and Allard, Baptiste and Alonso, Iv{\ifmmode\acute{a}\else\'{a}\fi}n and others},
	title = {{Terrestrial very-long-baseline atom interferometry: Workshop summary}},
	journal = {AVS Quantum Sci.},
	volume = {6},
	number = {2},
	year = {2024},
	month = jun,
	publisher = {AIP Publishing},
	doi = {10.1116/5.0185291}
}

@article{Ufrecht2020Nov,
	author = {Ufrecht, Christian and Di Pumpo, Fabio and Friedrich, Alexander and Roura, Albert and Schubert, Christian and Schlippert, Dennis and Rasel, Ernst M. and Schleich, Wolfgang P. and Giese, Enno},
	title = {{Atom-interferometric test of the universality of gravitational redshift and free fall}},
	journal = {Phys. Rev. Res.},
	volume = {2},
	number = {4},
	pages = {043240},
	year = {2020},
	month = nov,
	publisher = {American Physical Society},
	doi = {10.1103/PhysRevResearch.2.043240}
}

@article{DiPumpo2022Apr,
	author = {Di Pumpo, Fabio and Friedrich, Alexander and Geyer, Andreas and Ufrecht, Christian and Giese, Enno},
	title = {{Light propagation and atom interferometry in gravity and dilaton fields}},
	journal = {Phys. Rev. D},
	volume = {105},
	number = {8},
	pages = {084065},
	year = {2022},
	month = apr,
	publisher = {American Physical Society},
	doi = {10.1103/PhysRevD.105.084065}
}

@article{Abdalla2025,
author={Abdalla, Adam
and Abe, Mahiro
and Abend, Sven
and Abidi, Mouine
and others},
title={Terrestrial Very-Long-Baseline Atom Interferometry: summary of the second workshop},
journal={EPJ Quantum Technology},
year={2025},
month={Apr},
day={03},
volume={12},
number={1},
pages={42},
issn={2196-0763},
doi={10.1140/epjqt/s40507-025-00344-3},
url={https://doi.org/10.1140/epjqt/s40507-025-00344-3}
}

\end{document}